\documentclass[preprint]{aastex}
\usepackage{epsf}
\usepackage{emulateapj5}
\usepackage{longtable}
\usepackage{lscape}

\usepackage[pdfpagemode={UseOutlines},bookmarks=true,bookmarksopen=true,
   bookmarksopenlevel=0,bookmarksnumbered=true,hypertexnames=false,
   colorlinks,linkcolor={blue},citecolor={blue},urlcolor={red},
   pdfstartview={FitV},unicode,breaklinks=true]{hyperref}
  
\newcommand{\nstars}{238}
\newcommand{\nanc}{36}
\newcommand{\nnonanc}{202}
\newcommand{\nspectra}{1082}

\newcommand{\nnewbe}{128}

\newcommand{\ha}{H$\alpha$}
\newcommand{\hi}{H~{\sc i}}
\newcommand{\oi}{O~{\sc i}}
\newcommand{\nitrogeni}{N~{\sc i}}
\newcommand{\cli}{Cl~{\sc i}}
\newcommand{\sii}{Si~{\sc i}}
\newcommand{\siii}{Si~{\sc ii}}
\newcommand{\fei}{Fe~{\sc i}}
\newcommand{\feii}{Fe~{\sc ii}}
\newcommand{\caii}{Ca~{\sc ii}}
\newcommand{\ci}{C~{\sc i}}
\newcommand{\mgi}{Mg~{\sc i}}
\newcommand{\mgii}{Mg~{\sc ii}}
\newcommand{\hei}{He~{\sc i}}

\newcommand{\kms}{$\rm km\,s^{-1}$}
\newcommand{\vp}{${\Delta}v_{\rm p}$}

\newcommand{\rp}{$r_{\rm d}$}

\newcommand{\rs}{$R_{\rm *}$}
\newcommand{\vsini}{$v\,{\rm sin}\,i$}

\shorttitle{High-resolution, $H$-band spectroscopy of Be stars with SDSS-III/APOGEE}
\shortauthors{Chojnowski et al.}

\begin{document}

\title{High-resolution, $H$-band Spectroscopy of Be Stars with SDSS-III/APOGEE: I. New~Be~Stars,~Line~Identifications,~and~Line~Profiles}

\author{S. Drew Chojnowski\altaffilmark{1,2}}
\affil{Department of Astronomy, University of Virginia, P.O. Box 400325, Charlottesville, VA 22904-4325, USA}
\email{drewski@virginia.edu}

\author{David~G.~Whelan\altaffilmark{3}, John~P.~Wisniewski\altaffilmark{4}, Steven~R.~Majewski\altaffilmark{1}, Matthew~Hall\altaffilmark{1}, Matthew~Shetrone\altaffilmark{5}, Rachael~Beaton\altaffilmark{1}, Adam~Burton\altaffilmark{1}, Guillermo~Damke\altaffilmark{1}, Steve~Eikenberry\altaffilmark{6}, Sten~Hasselquist\altaffilmark{2}, Jon~A.~Holtzman\altaffilmark{2}, Szabolcs~M{\'e}sz{\'a}ros\altaffilmark{7}, David~Nidever\altaffilmark{8}, Donald~P.~Schneider\altaffilmark{9,10}, John~Wilson\altaffilmark{1}, Gail~Zasowski\altaffilmark{11}, Dmitry~Bizyaev\altaffilmark{2}, Howard~Brewington\altaffilmark{2}, J.~Brinkmann\altaffilmark{2}, Garrett~Ebelke\altaffilmark{2}, Peter~M.~Frinchaboy\altaffilmark{12}, Karen~Kinemuchi\altaffilmark{2}, Elena~Malanushenko\altaffilmark{2}, Viktor~Malanushenko\altaffilmark{2}, Moses~Marchante\altaffilmark{2}, Daniel~Oravetz\altaffilmark{2}, Kaike~Pan\altaffilmark{2}, Audrey~Simmons\altaffilmark{2}}

\altaffiltext{1}{Department of Astronomy, University of Virginia, P.O. Box 400325, Charlottesville, VA 22904-4325, USA}
\altaffiltext{2}{Apache Point Observatory and New Mexico State University, P.O. Box 59, Sunspot, NM, 88349-0059, USA}
\altaffiltext{3}{Department of Physics, Austin College, 900 N. Grand Ave., Sherman, TX 75090, USA}
\altaffiltext{4}{Department of Physics \& Astronomy, The University of Oklahoma, 440 W. Brooks St. Norman, OK 73019, USA}
\altaffiltext{5}{Department of Astronomy, The University of Texas at Austin, 2515 Speedway, Stop C1400 Austin, Texas 78712-1205, USA}
\altaffiltext{6}{Department of Astronomy, University of Florida, 211 Bryant Space Science Center, Gainesville, FL 32611-2055, USA}
\altaffiltext{7}{Department of Astronomy, Indiana University, Bloomington, IN 47405, USA}
\altaffiltext{8}{Department of Astronomy, University of Michigan, 830 Dennison 500 Church St. Ann Arbor, MI  48109-1042, USA}
\altaffiltext{9}{Department of Astronomy and Astrophysics, The Pennsylvania State University, University Park, PA 16802, USA}
\altaffiltext{10}{Institute for Gravitation and the Cosmos, The Pennsylvania State University, University Park, PA 16802, USA}
\altaffiltext{11}{Department of Physics \& Astronomy, Johns Hopkins University, Bloomberg Center for Physics and Astronomy Room 366, 3400 N. Charles Street, Baltimore, MD 21218, USA}
\altaffiltext{12}{Department of Physics and Astronomy, Texas Christian University, Box 298840, Fort Worth, TX 76129, USA}

\begin{abstract}
The Apache Point Galactic Evolution Experiment (APOGEE) has amassed the largest ever collection of multi-epoch, high-resolution ($R\sim22,500$), $H$-band spectra for B-type emission line (Be) stars. These stars were targeted by APOGEE as telluric standard stars and subsequently identified via visual inspection as Be stars based on {\hi} Brackett series emission or shell absorption in addition to otherwise smooth continua and occasionally non-hydrogen emission features. The {\nnewbe}/{\nstars} APOGEE Be stars for which emission had never previously been reported serve to increase the total number of known Be stars by $\sim6$\%. Because the $H$-band is relatively unexplored compared to other wavelength regimes, we focus here on identification of the $H$-band lines and analysis of the emission peak velocity separations ({\vp}) and emission peak intensity ratios (V/R) of the usually double-peaked {\hi} and non-hydrogen emission lines. {\hi} Br11 emission is found to preferentially form in the circumstellar disks at an average distance of $\sim$2.2 stellar radii. Increasing {\vp} toward the weaker Br12--Br20 lines suggests these lines are formed interior to Br11. By contrast, the observed IR {\feii} emission lines present evidence of having significantly larger formation radii; distinctive phase lags between IR {\feii} and {\hi} Brackett emission lines further supports that these species arise from different radii in Be disks. Several emission lines have been identified for the first time including {\ci}~16895, a prominent feature in the spectra for almost a fifth of the sample and, as inferred from relatively large {\vp} compared to the Br11--Br20, a tracer of the inner regions of Be disks. Emission lines at 15760~{\AA} and 16781~{\AA} remain unidentified, but usually appear along with and always have similar line profile morphology to {\feii}~16878. Unlike the typical metallic lines observed for Be stars in the optical, the $H$-band metallic lines, such as {\feii}~16878, never exhibit any evidence of shell absorption, even when the {\hi} lines are clearly shell-dominated. The first known example of a quasi-triple-peaked Br11 line profile is reported for HD~253659, one of several stars exhibiting intra- and/or extra-species V/R and radial velocity variation within individual spectra. Br11 profiles are presented for all discussed stars, as are full APOGEE spectra for a portion of the sample.


\end{abstract}



\section{Introduction} \label{intro}
Since the first observational description \citep{1931ApJ....73...94S} of the characteristic double-peaked emission lines of classical Be stars, a wealth of research has demonstrated that the emission lines are formed in geometrically-thin, equatorial circumstellar disks fed by gas ejected from the surfaces of rapidly rotating B stars \citep{2003PASP..115.1153P, 2013A&ARv..21...69}. Rapid rotation is certainly involved in the formation of these disks, but a comprehensive model of Be disk formation has yet to be created and efforts toward one are complicated by factors including the lack of examples of critically-rotating Be stars, star-specific peculiarities, and the requirement of an `on/off' switch to the Be phenomenon. For an uncertain but non-negligible fraction of Be stars the disks are transient, appearing in one epoch but not another \citep{2009ApJ...700.1216M, 2010ApJ...709.1306W}. A variable rotation speed ({\vsini}) that occasionally reaches or exceeds the critical breakup limit is an attractive concept for such a phenomenon \citep{2013A&A...559L...4R} that needs to be explored for a sample of transient Be stars. Non-radial pulsation and turbulence due to small-scale magnetic fields remain the most likely mechanisms, along with rapid rotation, responsible for the creation of the disks \citep{2013A&ARv..21...69}. When the disks are present, they appear to undergo Keplerian rotation \citep{2007A&A...464...59M, 2012ASPC..464..205W}, and many of the observational signatures are consistent with those predicted by the viscous decretion disk model \citep{1991MNRAS.250..432L, 2009A&A...504..915C, 2011IAUS..272..325C}. 

Multi-wavelength studies of Be disks are particularly valuable for diagnosing their structure because emission at different wavelengths originates from different physical locations within the disks \citep{2011IAUS..272..325C}. However, unlike in the optical wavelength regime where they have been extensively studied at high spectral and temporal resolution, only a limited number of Be star surveys have been performed at near-infrared (NIR) wavelengths, and these have typically utilized low spectral resolution \citep{2001A&A...371..643S} and small sample sizes \citep{1994A&A...284L..27M, 2000A&AS..141...65C, 2009PASP..121..125M, 2010AJ....139.1983G}. Detailed NIR spectroscopic studies of individual Be stars are more common \citep[e.g.,][]{2000A&A...355..187H, 2012MNRAS.423.2486M} and have been used to better diagnose the gas distribution within Be disks, including the structure of one-armed density waves \citep{2007ApJ...656L..21W, 2009A&A...504..929S, 2009A&A...504..915C}.

The Apache Point Observatory Galactic Evolution Experiment \citep[APOGEE;][]{2012AAS...21920506M} is actively providing the first ever bulk view of the high-resolution $H$-band properties and variability of Be stars. APOGEE is one of four surveys comprising the Sloan Digital Sky Survey III \citep[SDSS-III;][]{2011AJ....142...72E}. While the primary goal of the APOGEE survey is to measure the dynamical and chemical history of the Milky Way Galaxy using high-resolution $H$-band spectroscopic observations of $10^{5}$ red giant branch (RGB) stars, APOGEE devotes 35 fibers per 300-fiber pointing to observe hot stars as telluric standards. This, in addition to the surveys large sky coverage and multi-epoch observing strategy, has made APOGEE ideal for serendipitous Be discoveries and high-resolution NIR time series data of Be stars.

Here, we present the first catalog APOGEE Be (ABE) stars. An overview of the APOGEE survey and APOGEE data is provided in Section~\ref{apogeeoverview}, the Be sample is described in Section~\ref{besample}, and the identifications of observed metallic emission lines are discussed in Section~\ref{metallic}. Sections~\ref{peaksep} and \ref{vrratio} focus on quantitative and comparative analysis of emission double-peak separation ({\vp}) and double-peak intensity ratios (V/R). Commentary on the more unusual or exceptional Be stars within the ABE sample is interspersed throughout, and an atlas of Br11 profiles is provided in Section~\ref{lineprofs}. The appendix includes supplemental figures displaying full APOGEE spectra for stars with strong emission features, as well as an expanded stellar data table. Future work will focus on the observed spectral variability of sources with multi-epoch APOGEE data as well as follow-up optical spectroscopy for a subset of the sample.

\section{APOGEE overview} \label{apogeeoverview}

\subsection{APOGEE instrument and observations} \label{instrument}
The APOGEE instrument is a 300-fiber, $R\sim22,500$ spectrograph \citep{2010SPIE.7735E..46W} attached to the SDSS 2.5-meter telescope \citep{2006AJ....131.2332G} at Apache Point Observatory. APOGEE records a vacuum wavelength range of 15145--16955~{\AA} via an arrangement of three Teledyne H2RG $2048\times2048$ detectors. The detector layout consists of ``blue'', ``green'', and ``red'' detectors which cover 15145--15808~{\AA}, 15858--16433~{\AA}, and 16474--16955~{\AA} respectively, resulting in coverage gaps between 15808--15858~{\AA} and 16433--16474~{\AA}. Dispersion varies with wavelength, but the central dispersions of the blue, green, and red detectors are 0.326~{\AA} pix$^{\rm -1}$, 0.283~{\AA} pix$^{\rm -1}$, and 0.236~{\AA} pix$^{\rm -1}$ respectively. As with the original SDSS spectrograph \citep{2013AJ....146...32S}, APOGEE fibers are plugged into custom, pre-drilled aluminum plates which are loaded into the telescope's focal plane and which can cover $3^{\circ}$ diameter areas of sky. Each fiber has a 2{\arcsec} field of view.

The APOGEE survey uses the Two Micron All Sky Survey \citep[2MASS;][]{2006AJ....131.1163S} as a source catalog and focuses on observations of known or photometrically-likely RGB stars for its main science objective (230/300 fibers per plate). For calibration purposes, blank sky (35/300 fibers) and blue telluric standards (35/300 fibers) are also observed. Typical exposure times are 1-hr, and the number of repeat observations per field is approximately equal to the number of 1-hr observations needed to reach a combined signal-to-noise ratio per raw pixel (SNR) of 100 for stars at the field-specific $H$ magnitude limit. The bright limit for science targets is always $H=7.0$, while the faint limit is variable and can be $H=11.0$ (1-hr visit), $H=12.2$ (3 1-hr visits), $H=12.8$ (6 1-hr visits), $H=13.3$ (12 1-hr visits), or $H=13.8$ (24 1-hr visits). Cohorts of RGB targets with similar $H$ magnitudes are exchanged in and out of the observing sequence as SNR $ \sim 100$ is reached.

\subsection{APOGEE telluric standard stars} \label{aptell}
Hot O- and B-type (OB) stars are ideal candidates for telluric standard stars (TSS) in the $H$-band because the associated spectra are relatively featureless \citep{1998ApJ...508..397M, 2001A&A...371..643S}. Selection of the TSS for each APOGEE field is based on $H$ magnitude and non-reddening-corrected color rather than on intrinsic spectral properties \citep{2013AJ....146...81Z}, such that the TSS for a given field are simply the apparent bluest available stars. Thus, APOGEE makes no distinction between ``normal'' and emission-line stars other than to prevent from selection as TSS any stars that are reddened with respect to other stars (e.g. dusty B[e] stars) in the $3^{\circ}$ fields. 

Unlike the RGB science targets, TSS are restricted to $5.5\le H \le11.0$ and are therefore always expected to reach $\rm SNR\,\geq100$ in the typical hour exposures. In addition, the TSS for each APOGEE field are generally `locked-in,' meaning that they are observed every time their respective fields are observed rather than being traded in and out of the sequence as are the RGB stars. A more comprehensive description of TSS selection is presented in \citet{2013AJ....146...81Z}.

\subsection{APOGEE spectra} \label{apspec}
There are several details worth noting about the APOGEE spectra. Vacuum wavelengths given in angstroms ({\AA}) are used in APOGEE data and throughout this paper. Some of the spectra were recorded during APOGEE commissioning, prior to the instrument having achieved optimal focus. The resolution of red detector data in APOGEE spectra taken prior to $\rm MJD\,{\textless}\,55804$ (September 2011) is $R\sim16,000$, while the resolution is $R\sim22,500$ for all blue and green detector data regardless of date and for all post-55804 red detector data. The raw data is processed by an automated reduction pipeline (Nidever, in preparation) that extracts the spectra, performs flat-field and wavelength calibration, and performs sky and telluric corrections. APOGEE's reduction pipeline is designed to use sky and TSS exposures (see Section~\ref{aptell}) to remove airglow and telluric absorption lines from the high-resolution spectra. Because the airglow removal process has not yet been perfected, residuals from partially subtracted airglow lines remain in the final reduced data products. 

Since the APOGEE survey focuses on chemical abundance and radial velocity analysis, flux standard stars are not observed and therefore the spectra are not flux-calibrated. All spectra displayed in this paper were continuum normalized using the CONTINUUM task in IRAF by fitting low-order splines to sections of blank continuum adjacent to {\hi} Brackett lines, separately for each detector. Quoted emission line intensities and V/R intensity ratios refer to intensity relative to normalized continuum level ($F_{\lambda}/F_{\rm c}$). Due to the proximity of the Br12 and Br14 lines to the coverage limits of the green detector (21 and 27~{\AA} respectively), it was at times difficult to achieve a continuum fit that did not result in obviously incorrect intensity levels for those with respect to the other Brackett lines. The tendency of the full-width at continuum level for Brackett series emission lines to well exceed 1000~{\kms} was a further complication in salvaging the Br12 and Br14 lines, which are are de-emphasized from analysis for these reasons.

\begin{figure*}[htb!]
\includegraphics[width=18cm]{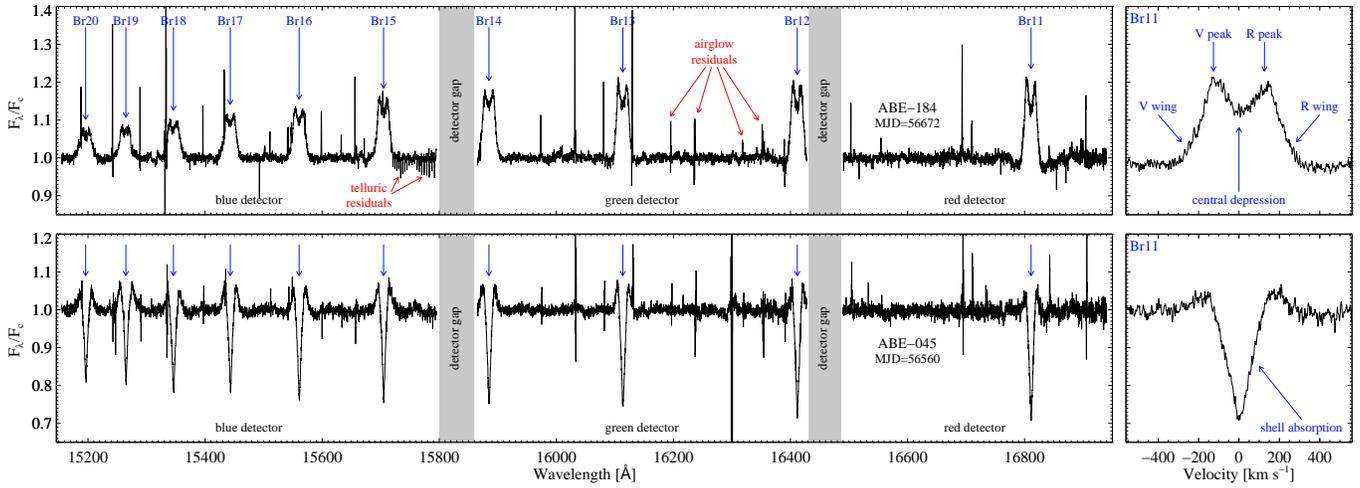}
\caption{Two examples of APOGEE spectra of Be stars. The `red', `green', and `blue' detectors and the gaps between detectors are labeled in the left panels and examples of residuals from the airglow and telluric removal process airglow residuals are labeled in the upper left panel (red arrows and text). The APOGEE Be star designations (ABE-184 and ABE-045) are shown along with the modified Julian date (MJD) of observation, and the Brackett series lines are labeled (blue arrows and text). Right-hand panels show the Br11 profiles on a velocity scale: ABE-184 exhibits a typical Be star line profile, where the central depression is a consequence of the disk geometry \citep[e.g.][]{1972ApJ...171..549H}, while the Brackett lines for ABE-045 are dominated by shell absorption resulting from occultation of the star by the disk in a nearly or exactly edge-on system \citep[e.g.][]{2006A&A...459..137R}. A color version of this figure is available online. \label{fig_intro}}
\end{figure*}

Figure~\ref{fig_intro} displays examples of APOGEE spectra for two newly-discovered Be stars, demonstrating the three-detector arrangement and associated coverage gaps. The APOGEE Be star IDs and modified Julian dates (MJD) of observation are provided above or below `red detector' continuum level, and commonly-observed emission lines (see Section~\ref{emlines}) are labeled with blue text and arrows. Examples of airglow residuals are noted with red text and arrows. The right-hand panels show Br11 line profiles from the same spectra on a velocity scale, with the line profile features of interest labeled. In most cases, Br11 is the strongest hydrogen line covered as well as the hydrogen line least likely to be affected by airglow or telluric contaminants should those be a significant issue. In subsequent figures, narrow contaminants (airglow and hot/cold/bad pixels) have been carefully trimmed from the spectra so as to avoid distraction from the features of astrophysical importance.

\subsection{Public availability of the spectra}
Both proprietary and publicly-available spectra are used and displayed in this paper. The publicly-available spectra were included in SDSS data release 10 (DR10: pertains to APOGEE data taken prior to MJD=56112), and the full data set will be made publicly-available in SDSS data release 12 (DR12: scheduled for December 2014). Shortly after DR12, we intend to convert the APOGEE Be star spectra to the format accepted by the Be Star Spectra Database \citep[BeSS;][]{2011AJ....142..149N} and deposit them there, ensuring convenient public access. More details on DR10-released APOGEE data can be found on the SDSS-III website (https://www.sdss3.org/dr10/irspec/).

\section{The ABE sample} \label{besample}

\subsection{Sample description} \label{sampledescription}
The sample at hand consists of {\nstars} Be stars that have been observed by APOGEE a total of {\nspectra} times. Of the {\nstars} ABE stars, {\nnonanc} were identified through periodic visual inspection of APOGEE spectra and {\nanc} were targeted intentionally to expand the subset of previously-known Be stars. 

We measured the velocity separations ({\vp}) of the violet (V) and red (R) emission peaks of all lines with well-defined peaks in all spectra of sufficient quality (typically $\rm SNR\,\textgreater50$; dependent on emission strength)  using the SPLOT feature of IRAF. Measurements pertaining to emission peaks coincident in wavelength position with strong airglow lines or diffuse interstellar bands (see Section~\ref{dib}) were thrown out. No attempt was made to remove underlying photospheric absorption prior to {\vp} measurement.

Table 1 provides the ABE identifiers, star names, 2MASS $H$ magnitude \citep{2003yCat.2246....0C}, literature spectral types and references where available (see Section~\ref{spectype}), and the mean {\vp} for the Br11 line from all APOGEE spectra for each source. Star names beginning with ``J'' are 2MASS designations, and newly-identified Be stars are indicated by bold font for the ABE ID. If a {\vp} measurement for the Br11 line could not be made in any of the available spectra despite evidence of Br11 emission or shell absorption, one of the following abbreviations is provided in place of a {\vp} value: ``\textbf{w}'' weak emission--peaks not discernible; ``\textbf{sp}'' single-peaked emission; ``\textbf{sh}'' shell absorption without resolved adjacent emission peaks; ``\textbf{as}'' severe asymmetry in emission peak heights such that only one peak is discernible (not the same as single-peaked); ``\textbf{bl}'' V peak of Br11 is severely blended with {\feii}~16792 (ABE-013); ``\textbf{tc}'' spectra are heavily contaminated by telluric features (ABE-058).

The ABE identifiers were assigned to avoid the use of sometimes lengthy survey identifiers which are the only star names available. Three groups of ABE stars are distinguished from one another by ABE ID as follows:
\begin{itemize}
   \item \textbf{ABE-001--ABE-202} refer to Be stars that were quasi-randomly targeted by APOGEE as TSS and subsequently identified as Be stars through visual inspection of the wavelength region encompassing Br11 and {\feii}~16878. To account for sources only producing emission lines in certain epochs, which was frequently the case, it was necessary to examine all ${\textgreater}70,000$ individual spectra for all ${\textgreater}17,000$ telluric stars. 
   \item \textbf{ABE-A01--ABE-A36} refer to Be stars that we targeted intentionally via internal proposals for APOGEE observations of ancillary (hence the `A' prefix of the ABE IDs) science targets falling within a subset of pre-planned APOGEE fields. Most of the intentionally-targeted Be stars are early-type (B3 and hotter) classical Be stars, showing stronger than average $H$-band emission in the APOGEE spectra, but two stars classified as B[e] in the literature were observed (ABE-A23 and ABE-A35) as was a reported Herbig Ae star (ABE-A33).
   \item \textbf{ABE-Q01--ABE-Q23} refer to stars which (a) had existing `emission line star' classifications in the literature, (b) appeared to be hot OBA stars in the APOGEE spectra, but (c) did not produce any discernible emission in any of the associated APOGEE data (all have multi-epoch data), or in others words, were $H$-band quiescent (hence the `Q' of the ABE IDs) during the observations.
\end{itemize}

\subsection{Literature spectral types} \label{spectype}
The {\hei}~17007 line, analogous to optical {\hei} in terms of utility as an effective temperature ($T_{\rm eff}$) diagnostic for OB stars \citep{1997AJ....113.1855B, 1998ApJ...508..397M} and the only non-hydrogen stellar absorption feature expected to be present for B-type stars (the earliest-O stars exhibit He~{\sc ii}~15723, 16923 absorption), is not covered by APOGEE spectra. Therefore, detailed spectral classification of OB stars is not possible with these data and the literature was perused for the existing spectral classifications included in Table~\ref{table_stars}. 

The Catalogue of Stellar Spectral Classification \citep[CSSC;][]{2013yCat....1.2023S} was the primary resource used for locating spectral type information, but some of the original sources of spectral types in the CSSC (primarily those pre-dating 1940) could not be tracked down. In those cases, the spectral types are enclosed in parentheses and the provided reference is ``C.'' Other second-hand spectral types, culled from modern compilations of historical data, are also enclosed in parentheses. The spectral types not enclosed in parentheses are therefore those that could be linked directly to the paper or catalog where the spectral type was determined or estimated.

\subsection{New Be star discoveries}
A total of {\nnewbe} Be stars have been identified as Be stars for the first time via Brackett series emission in APOGEE spectra. According to the Be Star Spectra Database \citep[BeSS;][]{2011AJ....142..149N}, which maintains a comprehensive database of classical Be and main sequence B[e] stars, 2070 Be stars are catalogued in the Milky Way and Magellanic clouds combined. The {\nnewbe} new Be stars presented in this work therefore represent a $\sim6.2\%$ increase in the number of known Be stars. 

\begin{figure*}[htb!]
\includegraphics[width=18cm]{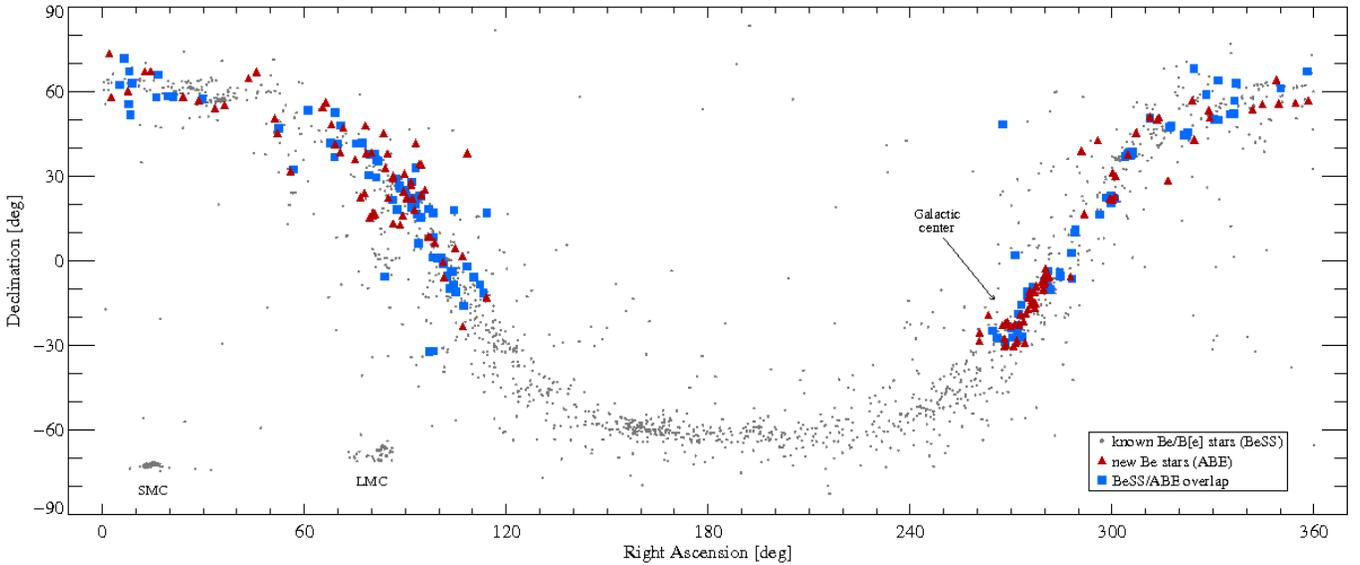}
\caption{This adaptation of \citet{2011AJ....142..149N} Figure 1 shows the RA and Dec positions of all the BeSS entries as black dots, known Be stars observed by APOGEE as squares (blue), and new Be stars discovered in the APOGEE survey as triangles (red). The Galactic Center and Magellanic Clouds are labeled. A color version of this figure is available online. \label{fig_radec}}
\end{figure*}

The positions of all {\nstars} ABE stars are shown in Figure~\ref{fig_radec}, along with the Be star entries included in the BeSS database \citep{2011AJ....142..149N}. Although APOGEE observes a large number of fields in the Galactic Halo, the majority (90\%) of Be stars observed during the survey reside along the plane of the Milky Way (at Galactic latitudes, {\textbar}$b${\textbar}~${\textless}\,10$), similar to the trend seen in BeSS.

\begin{figure*}[htb!]
\includegraphics[width=18cm]{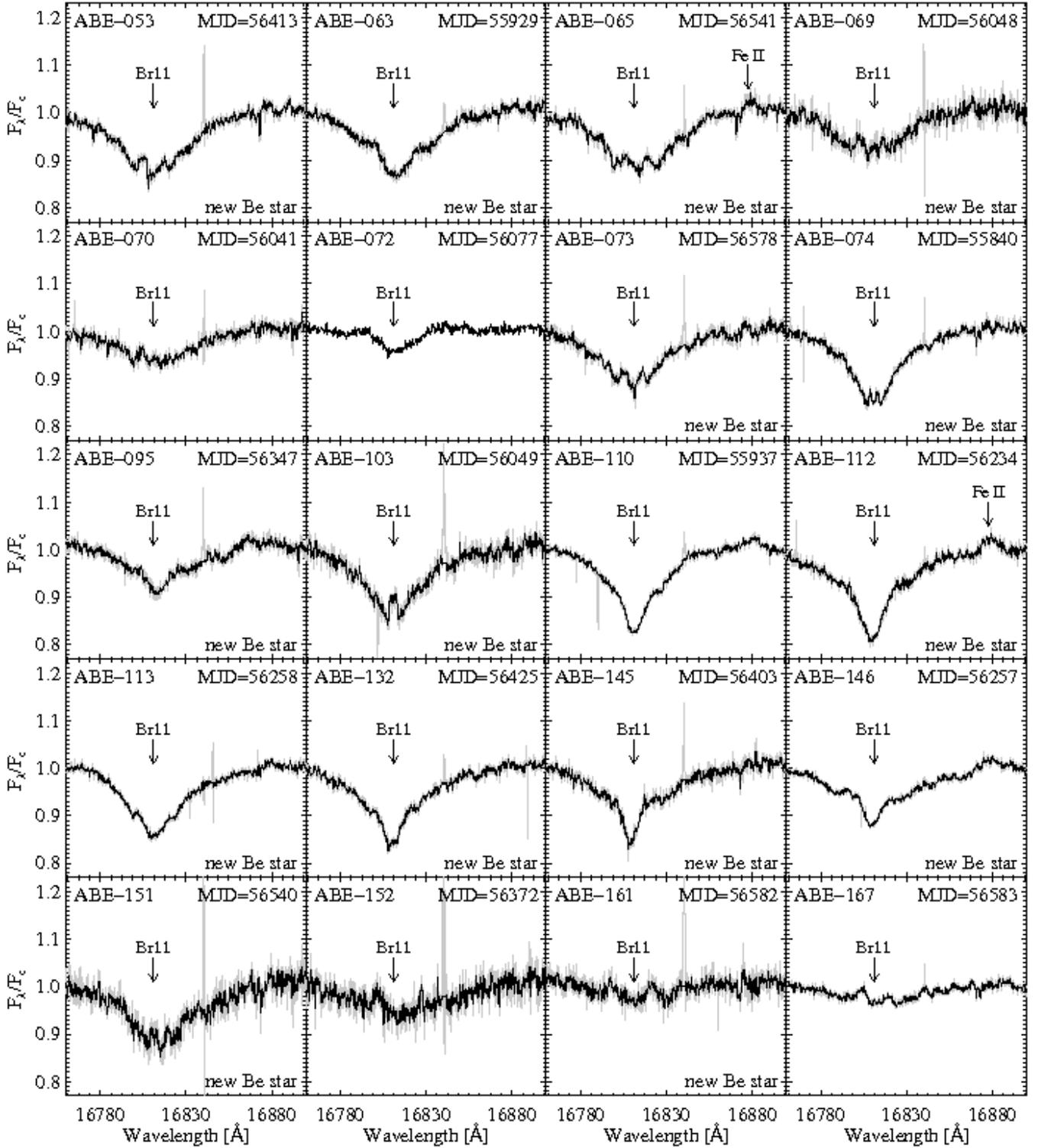}
\caption{The Br11 region for some new and previously-known Be stars showing very weak evidence of circumstellar emission. \label{fig_weakdisk}}
\end{figure*}

Stars included in the ABE sample were generally required to exhibit evidence of emission or shell absorption in at least the {\hi} Br11 line. The exceptions to this rule are ABE-111, ABE-196, and ABE-A06; these stars appear clearly to have emission from {\feii}~16878 (see Section~\ref{fe2lines}) despite very weak or no emission in the Brackett lines. Figure~\ref{fig_weakdisk} shows examples of new and previously-known Be stars representing the most borderline cases included the ABE sample. For many of these stars, the Br11 emission is sufficiently weak that double-peaks are not discernible. Rather, the photospheric Br11 absorption wings appear filled in with emission, creating `shoulders' on the line profiles (e.g. ABE-112) and making them easily distinguishable from purely photospheric lines profiles. {\feii} emission is also apparent for a number of these stars, despite weakness of the {\hi} emission.

The large number of new Be stars identified by APOGEE is due in large part to the high-resolution, high-SNR spectra which permit identification of very weak disk signatures (e.g. Figure~\ref{fig_weakdisk}) that might be overlooked in lower-resolution spectra or narrow-band photometry. Repeated observations of most of the stars ({\textgreater}1 observation for 93\% of sample) can provide confirmation of very weak disk signatures and also reveals transient Be disks, where Brackett series emission either fades away or appears unexpectedly from epoch to epoch (Wisniewski, in preparation). Among the reasons for 86/{\nnewbe} newly-identified ABE stars having been classified in the literature as normal O-, B-, or A-type stars is that the stars did not possess CS disks at the time the spectral types were determined or estimated. 

\subsection{ABE-144 and ABE-170: the brightest new Be stars}
The brightest newly-identified Be star among the ABE sample is ABE-170, $a.k.a.$ HR~2116 ($V=6.36$), and the second-brightest is ABE-144, $a.k.a.$ HD~189847 ($V=6.92$). The lack of a prior indicator of emission lines for ABE-144 may be due to few or no historical spectroscopic observations of the star beyond \citet{1962JO.....45..349F}. It is unclear whether or not Balmer series emission would have been noticed in that study. ABE-170 has been observed spectroscopically in more recent studies including \citet{2002ApJ...573..359A} and \citet{2005AJ....129..809S}, but the spectra used in both of those papers were limited in wavelength coverage to the region encompassing {\hei}~4471 and {\mgii}~4481, such that emission at H$\alpha$ or H$\beta$ would not have been recognized if present. Among the possible reasons for the Be nature of these stars not having been previously recognized is that ABE-144 and ABE-170 were normal B stars during past observations \citep[similar to, e.g.,][]{2000AAS...197.4623B}.

\subsection{Observed emission lines} \label{emlines}
The emission lines detected in APOGEE's wavelength range are listed in Table~\ref{table_vpsummary}. For each line, Table~\ref{table_vpsummary} provides the (1) line identification, (2) laboratory rest wavelength, (3) observed wavelength (see concluding paragraph of this section), (4) the difference between laboratory and observed wavelengths, (5) lower level energy $E_{i}$, (6) upper level energy $E_{k}$, (7) transition strength expressed as log($g_{i}f_{ik}$), (8) for metallic lines only, the number of confident and possible detections, (9) the number of stars for which {\vp} was measured, (10) the range of {\vp} measurements, (11) the average of all {\vp} measurements, (12) and other transitions possibly contributing to the observed emission line profiles.

Attempts to identify all non-hydrogen (metallic) lines, described in Section~\ref{metallic}, made use primarily of Peter van Hoof's Atomic Line List v2.05b16\footnote{http://www.pa.uky.edu/$\sim$peter/newpage/} (PLL from here on) and to a lesser extent the NIST Atomic Database \citep{Kramida13} and the Kurucz line list\footnote{1995 Atomic Line Data (R.L. Kurucz and B. Bell) Kurucz CD-ROM No. 23. Cambridge, Mass.: Smithsonian Astrophysical Observatory}. The identities of metallic emission lines at 15760~{\AA} and 16781~{\AA} remain ambiguous due to few transitions around the correct wavelengths having available transition probability data needed for confident identification. These lines, referred to as $\lambda15760$ and $\lambda16781$ throughout this paper, are discussed in more detail in Section~\ref{unidentified}. Since forbidden line emission was present for only one source, ABE-A23 $a.k.a.$ MWC~922, the central star of the Red Square Nebula \citep{2007Sci...316..247T}, Table~\ref{table_vpsummary} is limited to the permitted (E1) transitions observed for Be stars. The $H$-band spectrum of MWC~922 is sufficiently different from the rest of the sample and sufficiently more complex that an in-depth analysis is currently being pursued separately (Whelan, in preparation). 

The observed wavelengths as well as the differences between laboratory and observed wavelengths, provided in columns (4) and (5) of Table~\ref{table_vpsummary}, pertain to the average position of double emission peaks for each line plus a correction factor based on the Doppler shift found for the Br11 line. Br11 is the strongest line covered for these stars and provides the most reliable peak position measurements, so correction to rest frame was done simply by adding to the observed wavelength of each line the difference between Br11 emission peak midpoint and Br11 rest wavelength. 

\section{Non-hydrogen line identification} \label{metallic}

\subsection{DIBs} \label{dib}
\begin{figure*}[t!]
\includegraphics[width=18cm]{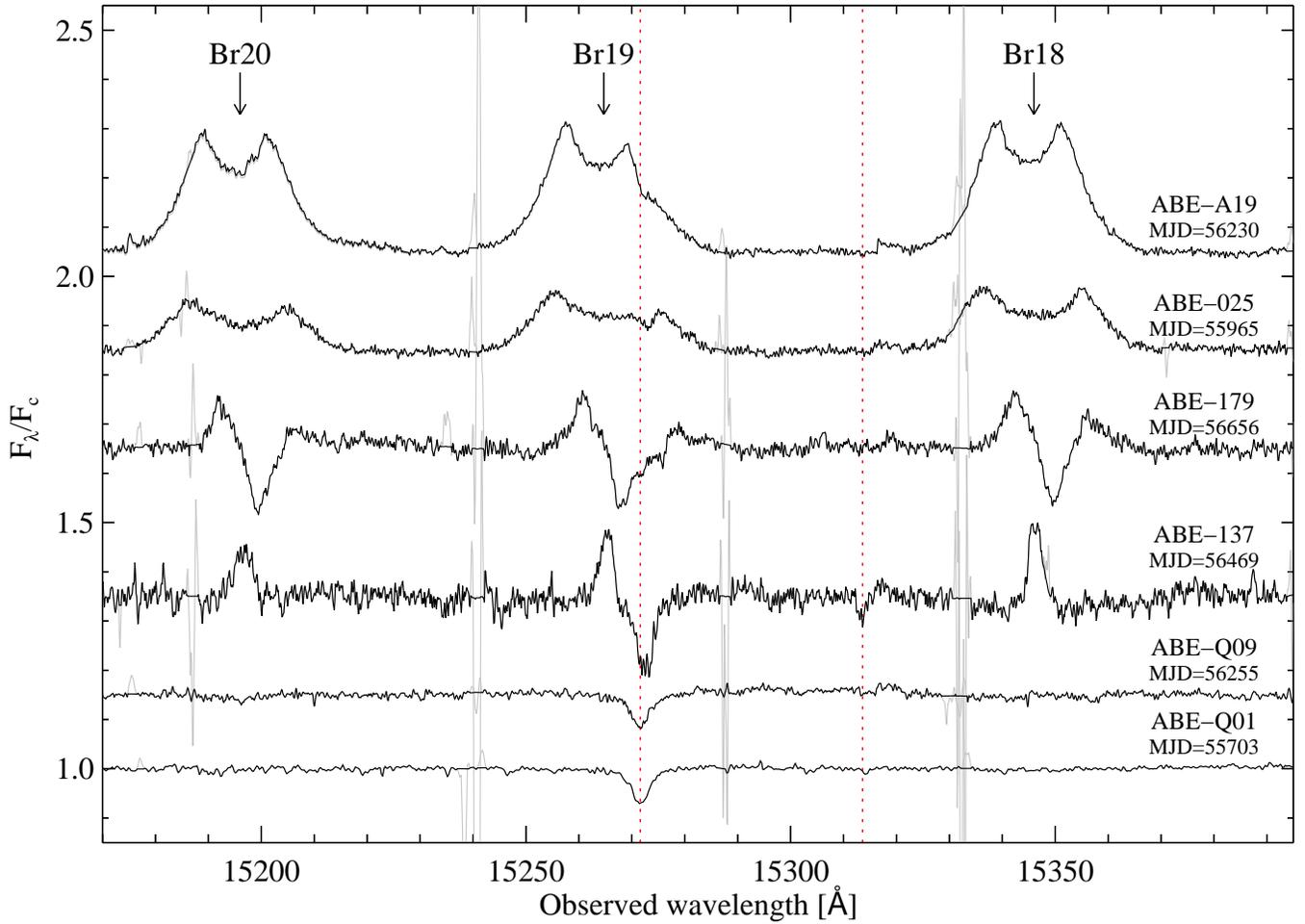}
\caption{Spectra for six stars with visible DIB~15271 absorption around or on the Br19 line. No correction for radial velocity has been applied to the spectra. The dotted lines (red) mark DIB~15271 and another likely DIB at $\sim15314$ that appears in numerous APOGEE spectra. A color version of this figure is available online. \label{fig_dib}}
\end{figure*}

The diffuse interstellar band (DIB) at 15271~{\AA}, discovered by \citet{2011Natur.479..200G}, is present in most of the ABE spectra and in numerous APOGEE spectra \citep{2014arXiv1406.1195Z}. Because DIB~15271 usually falls on or near the Br19 R emission peak, Br19 peak separation measurements are omitted from this paper. Examples of DIB 15271 absorption (marked with red dotted lines) in spectra for four active Be stars and two currently emission-less stars are displayed in Figure~\ref{fig_dib}. Other DIBs (15615, 15651, 15671~{\AA}) discussed in \citet{2011Natur.479..200G} are present for most objects with DIB 15271, as are other possible DIBs at $\sim$15314~{\AA} and $\sim$16154~{\AA}. Of the spectra shown in Figure~\ref{fig_dib}, DIB 15314 appears most prominently in the spectrum for ABE-137.

\subsection{{\feii}} \label{fe2lines}
The {\feii}~16878 line appears in emission for between 32--46\% (upper limit includes weak or ambiguous detections) of the 238 active Be stars discussed here, making it the most frequently-observed metallic feature in APOGEE's coverage of the $H$-band. For stars with very strong {\feii}~16878, the much weaker {\feii}~16792 also appears in emission but is usually blended with Br11. As for {\feii}~16878, proximity of the feature to {\ci}~16895 often leads to a blend of the two lines, especially since {\ci} emission is always broad compared to {\feii} (see Section~\ref{vpeak}).

\begin{figure*}[t!]
\includegraphics[width=12cm]{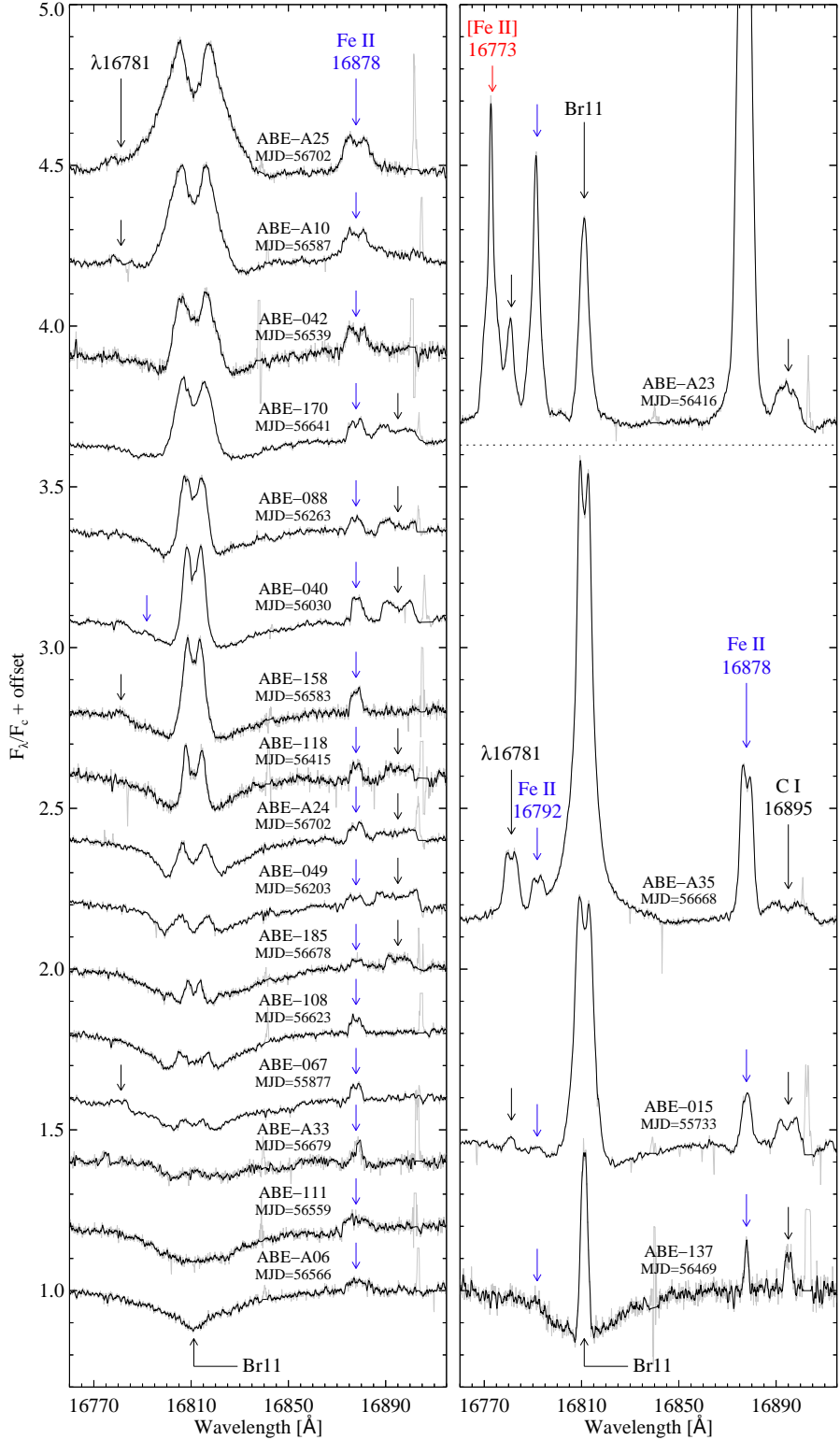}
\caption{Be star spectra with combinations of emission from $\lambda16781$, {\feii}~16792 and 16878 (blue), and {\ci}~16895 over a spectrum of Brackett series emission strength. The left and right panels show the same wavelength and intensity ranges. A dotted line separates the unclassified B[e] star ABE-A23 (MWC~922) from the other stars; ABE-A23 is unique among this sample (see Section~\ref{emlines}) in being the only source to show forbidden line emission (mostly [Fe~{\sc ii}]). A color version of this figure is available online. \label{fig_fe2panel}}
\end{figure*}

Examples of stars with emission from one or both $H$-band {\feii} lines are presented in Figure~\ref{fig_fe2panel}. The left panel demonstrates the wide range of {\hi} strength corresponding to {\feii} detections. As is seen quite clearly for the lower-most stars (ABE-A06, ABE-111) in the left panel of Figure~\ref{fig_fe2panel}, {\feii} emission may be present even when there is no perceptible emission from Brackett series lines, contrary to the finding of \citet{2001A&A...371..643S}. ABE-A06 has been a Be-shell star at various epochs (see BeSS spectra), but in the APOGEE data exhibits only very weak filling of the Br11 photospheric absorption wings in addition to the weak {\feii}~16878 emission that, for ABE-A06, persists in four spectra sparsely covering 0.77 years. Less evidence is available for {\hi} emission in the case of ABE-111, despite the {\feii} feature appearing in all six APOGEE covering 2.29 years. Though not shown in Figure~\ref{fig_fe2panel}, ABE-196 also lacks convincing evidence of {\hi} emission and yet exhibits {\feii}~16878 emission in all 13 APOGEE spectra covering 3.02 years. Line profile variability in the Brackett lines is observed for all three stars and is likely due to varying degrees of emission filling, but lack of knowledge of the true photospheric absorption profiles prevents us from confidently claiming {\hi} emission is present. 

The right panel of Figure~\ref{fig_fe2panel} focuses on some of the more extreme Be stars in this sample, starting with the obvious outlier ABE-A23, an unclassified B[e] star \citep{1998A&A...340..117L} for which the exceptionally strong {\feii} and [{\feii}] emission lines reported by \citet{1992ApJ...398..278R} dominate the APOGEE spectra. In contrast to ABE-A23, where the emission lines all appear truly single-peaked, ABE-137 is likely a classical Be star viewed at an inclination, $i$, of nearly or exactly zero. The Brackett series lines for ABE-137 show some peak structure even though the peaks are not resolved. The {\feii}~16878 line is very narrow and pointed but double peaks are resolved in the {\ci}~16895 line, suggesting the presence of a circumstellar disk. ABE-A35 exhibits strong {\feii} emission, and is the only source of the ABE sample for which the {\vp} of {\feii}~16792 could be measured. The average {\feii} peak separations from five spectra for ABE-A35 are in good agreement: {\vp}({\feii}~16878)~=~$52.6\pm2.53$~{\kms}, {\vp}({\feii}~16792)~=~$48.3\pm4.01$~{\kms}. The lower resolution of red detector data from APOGEE commissioning data is likely a factor in the single-peaked appearance of the {\feii} lines for ABE-015.

\subsection{{\feii} profiles as a function of inclination} \label{fe2shell}
In past studies, optical {\feii} lines have been used \citep{1996A&A...308..170H} to establish a strict definition of the shell, or edge-on, class of Be stars. Photospheric {\feii} absorption lines are usually observed at greatest strength for A-F supergiants \citep{2009ssc..book.....G}, so if the central depression of an {\feii} emission line for a Be star extends below undisturbed, adjacent continuum level, the implication is that the the disk is viewed at sufficiently large inclination that our line of sight passes through an appreciable volume of cool gas in the inner, equatorial disk. It is a well-known fact that {\feii} and Ti~{\sc ii} shell lines are among the strongest metallic features present in the spectra of edge-on Be stars.

\begin{figure*}[htb!]
\includegraphics[width=18cm]{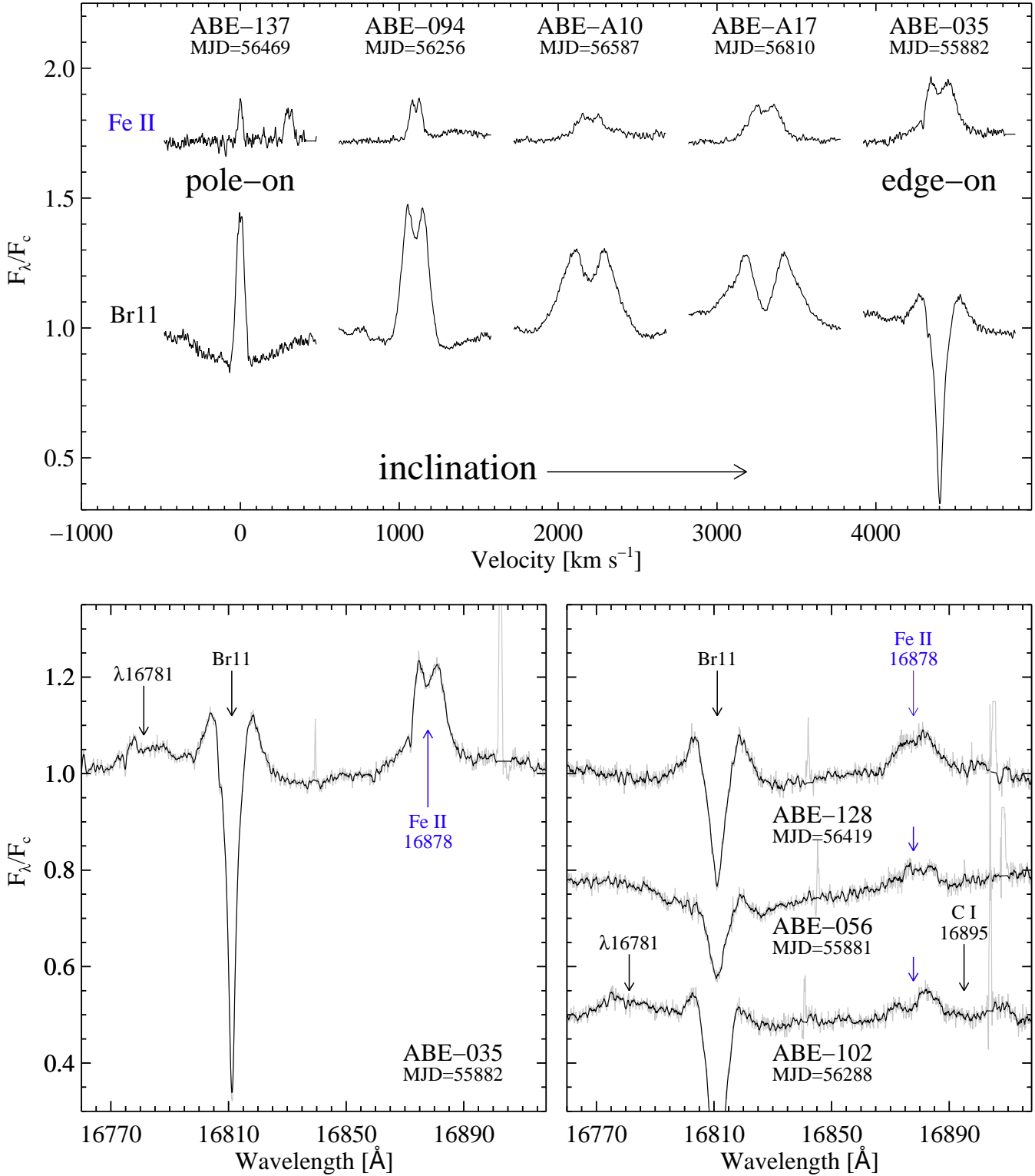}
\caption{(Top panel) An assortment of observed spectra, showing the variety of metallic and hydrogen line profiles observed in the sample as a function of inclination angle. (Left, bottom) A portion of a spectrum for ABE-035, highlighting the immunity of metallic emission lines to shell absorption, a fact that is observed in all shell absorption sources (examples shown on right, bottom). A color version of this figure is available online. \label{fig_fe2shell}}
\end{figure*}

In contrast to the observed behavior of optical {\feii} lines, stars with obvious shell absorption in the Brackett series lines exhibit no evidence of shell absorption in the $H$-band {\feii} lines nor in any of the covered metallic lines, such that the {\feii} line profile shapes for pole-on Be stars differ from those of edge-on Be stars only in line width. This fact is demonstrated in Figure~\ref{fig_fe2shell}, where the upper panel compares Br11 and {\feii} profiles for five stars viewed over a range of inclination angles. As can be seen, the {\feii} profiles are pure emission regardless of the what form the {\hi} profiles take. The lower right panel of Figure~\ref{fig_fe2shell} presents additional examples of {\hi}-shell stars with {\feii} emission, while the lower left panel (as well as the edge-on example in the upper panel) highlights ABE-035, the most extreme shell star within this sample in terms of Brackett series shell depth.


\subsection{{\ci} and other neutral lines} \label{c1lines}
An emission line at 16895~{\AA} is identified for the first time as {\ci}~16895.031 and is observed for between 18--26\% of the 238 Be stars. Figure~\ref{fig_fe2panel} displays 11 examples of stars with {\ci} emission and the strongest detections will be discussed in Section~\ref{abe084}. A {\ci}~16895 absorption line is present in numerous APOGEE spectra of A-F stars, but the line is typically not present for OB stars unless in emission. Except in the case of very narrow-lined Be stars (e.g. ABE-015, ABE-040 in Figure~\ref{fig_fe2panel}), the R peak of the {\ci}~16895 emission profile is frequently compromised by a strong airglow line around $\sim16904$~{\AA}.

Prior mentions in the literature of NIR {\ci} emission include \citet{2007A&A...465..993G} and \citet{2009A&A...504..929S}, where several {\ci} emission lines were detected around 10700~{\AA} in Be star spectra. Spectra showing {\ci}~16895 emission have been included in a number of papers, but the line is usually either confused and/or blended with {\feii}~16878, or not identified at all. \citet{2000ASPC..214..468A} noticed the {\ci}~16895 line in a subset of low-resolution Be star spectra and realized that it was probably not {\feii}~16878 due to the measured wavelength of the line ($\sim$16893~{\AA}, or $\sim15$~{\AA} from the {\feii} wavelength). \citet{2003A&A...408..313K} presented medium resolution $H$-band spectra of three young stellar object (YSO) candidates, one of which, IRAS~17441-2910, was found to be a very strong emission line source. A plot of the spectrum shows single-peaked Br11 and {\feii}~16878 emission and strong double-peaked {\ci}~16895 emission, but the authors did not comment on the latter. 

NIR emission from {\ci} is not limited to classical Be stars. {\ci}~16895 emission was present in a high-resolution spectrum shown by \citet{2012ApJ...752...11K} for the Herbig B[e] star V921~Scorpii, and the {\ci} emission line is also present in APOGEE spectra for both B[e] stars observed by APOGEE to date (see right-hand panel of Figure~\ref{fig_fe2panel}). Even luminous blue variable (LBV) stars display evidence of {\ci} emission lines \citep{2007A&A...465..993G}, suggesting that NIR {\ci} emission is ubiquitous across a wide range of evolutionary states.

\subsubsection{{\ci}--strong Be stars} \label{abe084}
\begin{figure*}[htb!]
\includegraphics[width=18cm]{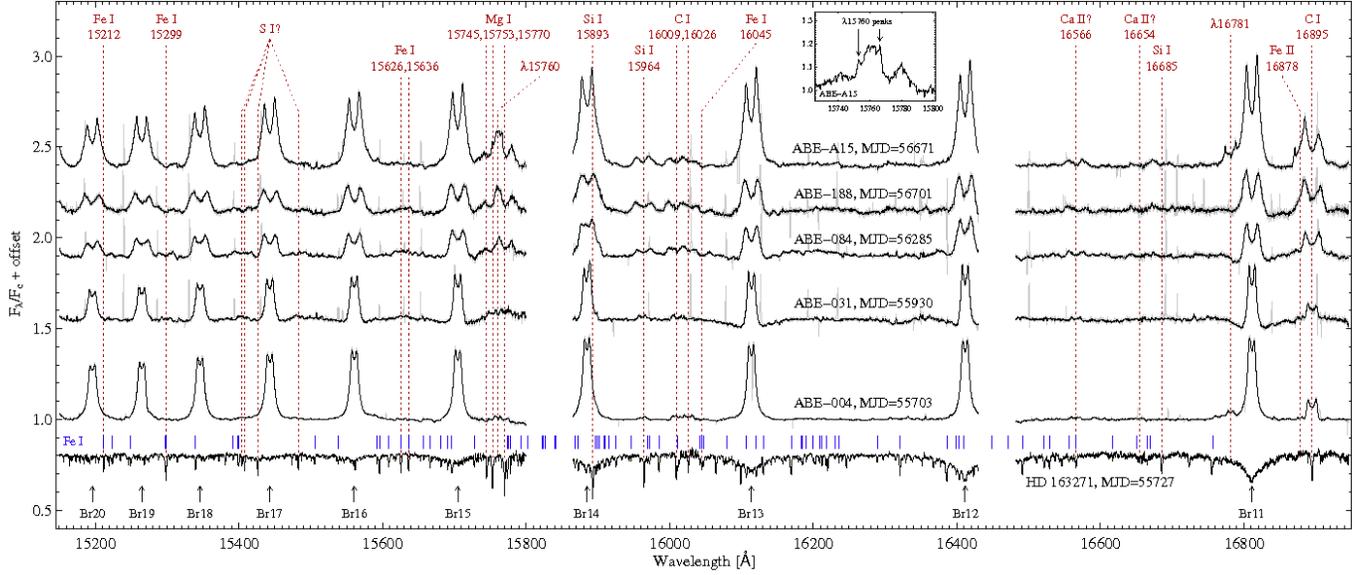}
\caption{Spectra of five Be stars (ABE-A15, ABE-188, ABE-084, ABE-031, ABE-004) with strong {\ci}~16895 emission and many weak, double-peaked metallic emission features. The spectrum of a strong-metal-lined A star (HD~163271) is included to demonstrate that the additional emission lines for these four Be stars correspond to absorption lines for cooler stars. The small lines (blue) above the A star spectrum mark the positions of numerous {\fei} lines with log($g_{i}f_{ik}$)~{\textgreater}~$-3$. A color version of this figure is available online. \label{fig_peculiar}}
\end{figure*}

Abnormally strong {\ci}~16895 emission is accompanied by weaker, similarly-profiled emission lines from neutral and singly-ionized species in the spectra for at least five ABE stars. Figure~\ref{fig_peculiar} displays full APOGEE spectra for ABE-A15, ABE-188, and ABE-084, ABE-031, and ABE-004, the best examples of this marked deviation from the typical $H$-band emission line content for Be stars. The {\ci}~16895 emission is blended with {\feii}~16878 for ABE-A15 and possibly also for ABE-188. Two other {\ci} lines at 16009.27~{\AA} and 16026.08~{\AA} are blended in emission for these stars, leading us to refer to the group as `{\ci}-strong' Be stars. 

Most of the metallic emission features for the {\ci}-strong Be stars correspond to strong absorption lines for late-A and cooler stars. An APOGEE spectrum for HD~163271, which is either a single metallic-line A star (A2/A3m) or the superposition (A2/A3+F0) of an A star with an F star \citep{1988mcts.book.....H, 1991A&AS...89..429R}, is provided in Figure~\ref{fig_peculiar} to demonstrate the typical line content for A-F stars. Small blue line segments indicate the numerous strong {\fei} lines covered, with {\fei}~15299 being the strongest and appearing in emission for the Be stars. It is likely that emission from other {\fei} lines is involved in much of the blending in the {\ci}-strong Be star spectra. 

Detections of resolved emission peaks for the S~{\sc i} lines labeled in Figure~\ref{fig_peculiar} are unavailable, but the lines may contribute to the weak bumps and blending around Br17. The transition probability measures for these S~{\sc i} lines suggests S~{\sc i}~15426 should be the strongest of them and the A star spectrum appears to confirm this. Since the Br17 profiles for the {\ci}-strong Be stars do not appear distorted by significant underlying emission from other lines however, it is unclear whether the S~{\sc i} lines are actually observed as emission features.

The strong emission lines redward of Br15 are due partly to several {\mgi} lines, with the strongest contributions being {\mgi}~15753.291 and {\mgi}~15770.149. The {\mgi} lines are also seen weakly in emission and unblended for ABE-149; all of the emission lines are single-peaked for that source, including {\hi}, {\feii}~16878, {\ci}~16895, and the {\mgi} lines (see Appendix A). Above the ABE-A15 spectrum in Figure~\ref{fig_peculiar} is a small panel that zooms in on the {\mgi} blend for ABE-A15, demonstrating that emission from $\lambda15760$ is also a major contributing factor in the blend. Black arrows in the small panel point out the sharp $\lambda15760$ peaks that mimic the sharp $\lambda16781$ peaks. ABE-004 similarly has the $\lambda15760$ and $\lambda16781$ lines clearly in emission. 

As for the line around 15964~{\AA}, PLL suggests two possible identities: {\cli}~15964.11 and {\sii}~15964.4218. Since other covered {\cli} lines are expected to be stronger than {\cli}~15964.11 are covered but do not appear in emission (e.g. {\cli}~15524.70), {\sii} seems the more likely to cause the 15964~{\AA} emission. The line blended with Br14 (most noticeable for ABE-084 and ABE-031) is suspected to be {\sii}~15892.7713, the next strongest {\sii} line covered after {\sii}~15964.4218. 

The weak double-peaked line around $\sim$16565~{\AA} is possibly {\caii}~16565.59, but the ambiguous detection of {\caii}~16654.43 calls the {\caii} identification into question since the latter line should be stronger. On the other hand, the position of {\caii}~16654.43 corresponds to a strong telluric band which is poorly-corrected and may cause the ambiguity. Emission from the {\caii} triplet (8498, 8542, 8662~{\AA}) is observed for some Be stars \citep{1947ApJ...105..212H, 1976IAUS...70...59P}, so $H$-band {\caii} emission would not be terribly unexpected. A {\ci} line at 16564.13~{\AA} probably does not contribute since similar {\ci} lines, covered and expected to be stronger than {\ci}~16564, fail to appear.

The cause of the strong {\ci}~16895 in addition to other weaker emission lines for the {\ci}-strong Be stars remains unknown. Based on the available examples however, such as ABE-031 where the weak emission features persist in 12 spectra covering 1.2 years, the phenomenon appears to be permanent rather than a particular stage of short- or medium-term intrinsic variability.

\subsection{$\lambda15760$ and $\lambda16781$} \label{unidentified}
\begin{figure*}[htb!]
\includegraphics[width=8cm]{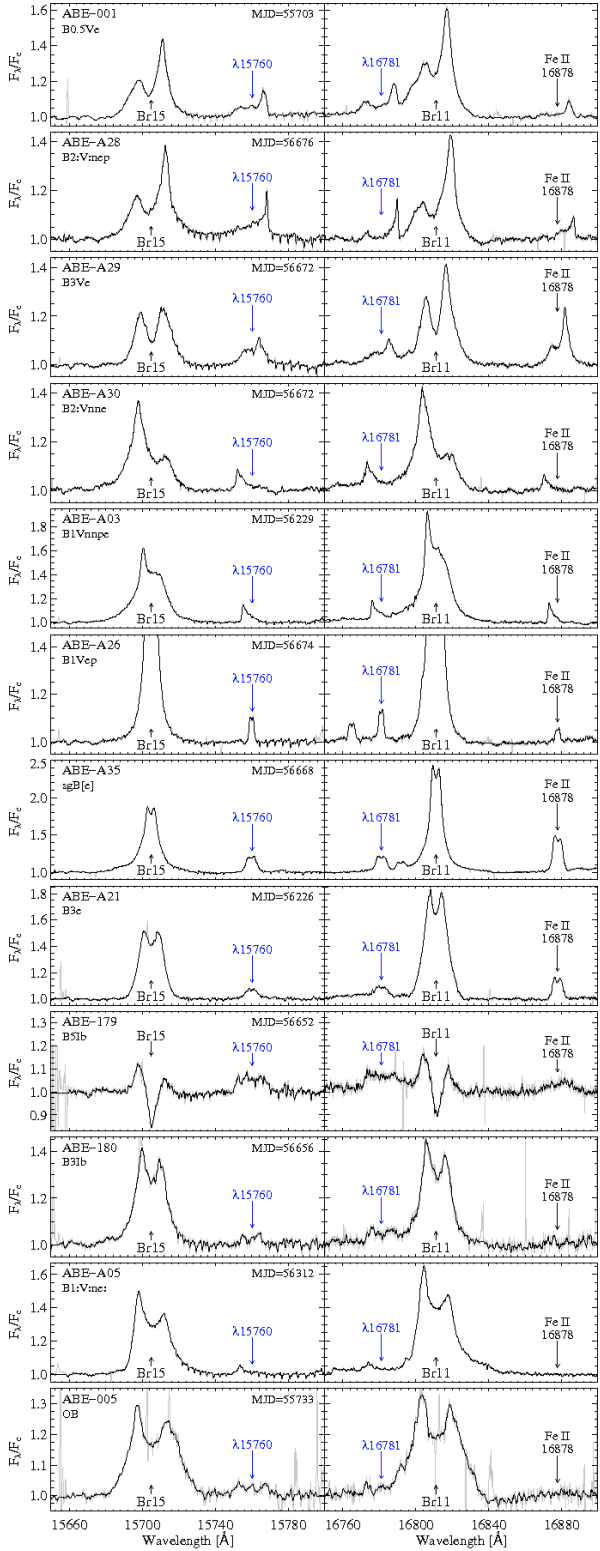}
\caption{The identifications of $\lambda15760$ and $\lambda16781$ are uncertain; however, these lines are never detected separately and in most cases {\feii}~16878 emission is also present. The three lines always share a common V/R orientation, but the {\feii} intensity varies with respect to $\lambda15760$ and $\lambda16781$. Small absorptions around the $\lambda15760$ line are telluric correction artifacts. A color version of this figure is available online. \label{fig_unidentified}}
\end{figure*}

The $\lambda15760$ and $\lambda16781$ emission lines discussed by \citet{2001A&A...371..643S} are present for between 15--21\%. As is demonstrated in Figure~\ref{fig_unidentified}, these lines always appear together with matching intensity and V/R orientation. In the available examples where peak separations were measurable for $\lambda15760$ and $\lambda16781$, those values are nearly identical as well (see Section~\ref{vpeak}). {\feii}~16878 is usually detected in unison with $\lambda15760$ and $\lambda16781$, but this is not a strict rule. Non-detection of {\feii}~16878 is accompanied by detections $\lambda15760$ and $\lambda16781$ for ABE-180, ABE-A05, and ABE-005, the three lower-most stars represented in Figure~\ref{fig_unidentified}. 

The $\lambda15760$ line has been identified as {\feii} in several past papers \citep{2001A&A...371..643S, 2001ApJ...551L.101S, 2012ApJ...752...11K}. In a study of $\eta$ Carinae, \citet{1994ApJ...436..292H} was apparently the first to note proximity of $\lambda15760$ to an {\feii} transition. The authors of that paper appended a question mark to the {\feii} identification listed in an emission line table, but it seems that over the years the question mark was forgotten. PLL lists an {\feii} line at 15761.78~{\AA}, but no indication of expected transition strength is available. NIST provides wavelength for different {\feii} transitions, at 15759.720~{\AA} and 15760.563~{\AA}, with the lower energy levels again more than doubling those of {\feii}~16878 (5.5 eV versus 13.4 eV) and again lacking transition strength indication. A firm identification for this emission line remains elusive.

Whereas \citet{2001A&A...371..643S} restricted the possible identifications for $\lambda16781$ to [{\feii}]~16773 and {\feii}~16792, the much higher-resolution APOGEE spectra rule out both of those lines as possibilities (see ABE-A23 spectrum in Figure~\ref{fig_fe2panel}). PLL includes several {\hei} lines around 16780~{\AA}, but considering that an absorption line is never seen at this wavelength for normal OB stars, $\lambda16781$ is probably not {\hei}. Also listed in PLL is an {\oi} multiplet at 16781.7~{\AA}, lacking transition probability data and being quickly ruled out by non-detection of other {\oi} lines covered and expected to be stronger. Through similar argument, other lines listed in PLL and NIST around $\lambda16781$ are readily ruled out as possibilities. 

Whatever the identities of $\lambda15760$ and $\lambda16781$, the features behave similarly to {\feii} in being present as emission lines or not present at all: no corresponding absorption for features are seen for APOGEE-observed stars of any type. Reliable spectral types have been reported for 24 ABE stars with $\lambda15760$ and $\lambda16781$ detections and 20/24 are B3 or hotter, so it is possible that $\lambda15760$ and $\lambda16781$ are relatively high-ionization lines. One possible example of $\lambda16781$ being detected despite absence of $\lambda15760$ is ABE-A36, a peculiar star discussed in Section~\ref{hphaselag}. However, the bump in the V wing of Br11 for ABE-A36 is not sufficiently convincing to cause us to doubt that $\lambda15760$ and $\lambda16781$ should always be expected to appear simultaneously.

\subsection{{\nitrogeni}}
\begin{figure*}[htb!]
\includegraphics[width=18cm]{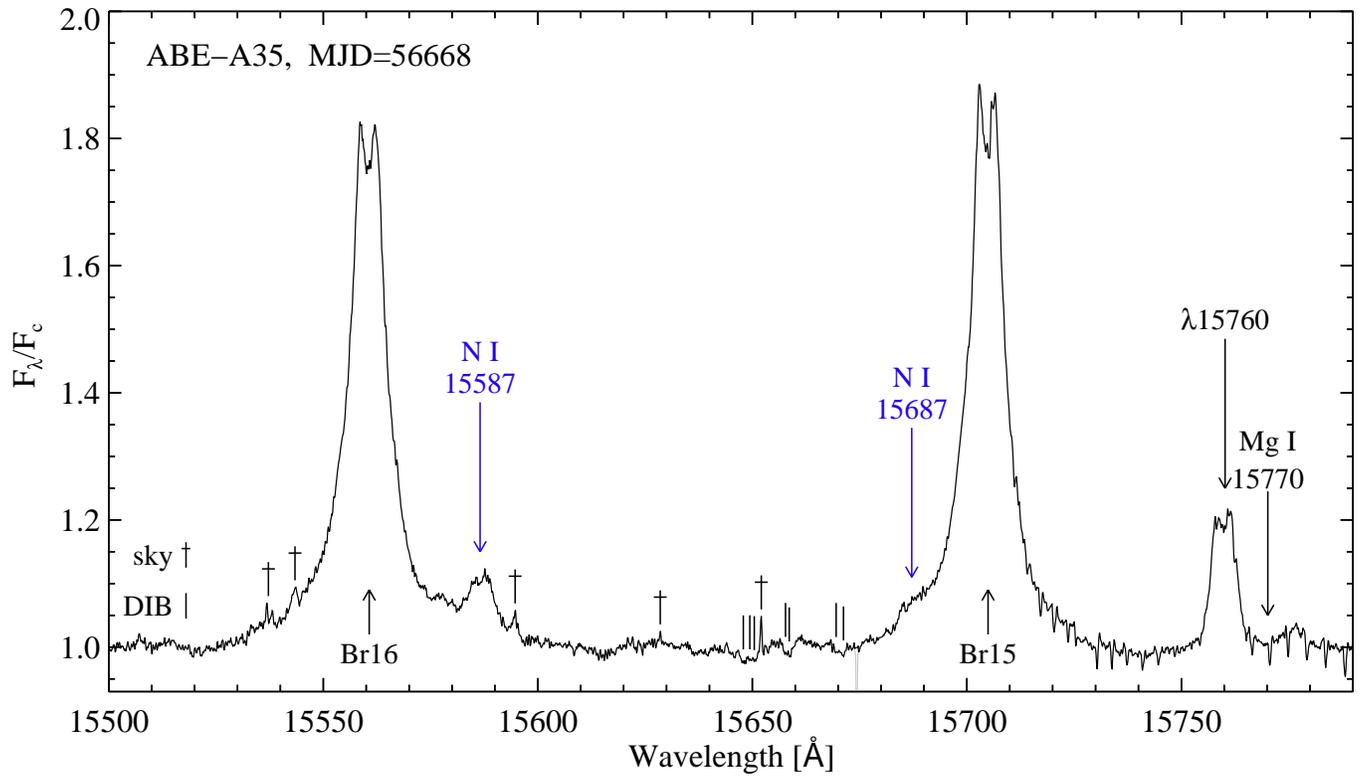}
\caption{Emission from {\nitrogeni} is seen only for ABE-A35, a supergiant B[e] star \citep{1998MNRAS.298..185E, 2013A&A...558A..17O}. Both lines, {\nitrogeni}~15587 and {\nitrogeni}~15687, are partially blended with {\hi} emission wings. A color version of this figure is available online.  \label{fig_ni15587}}
\end{figure*}

The expected strongest and second strongest {\nitrogeni} lines covered are seen in emission for ABE-A35. Figure~\ref{fig_ni15587} shows a portion of a spectrum for ABE-A35 encompassing the {\nitrogeni}~15586.545 and {\nitrogeni}~15687.160 lines as well as Br16, Br15 and $\lambda15760$. Neither of the {\nitrogeni} lines are detected for any other objects beyond ABE-A35, but they are present in all five APOGEE spectra for ABE-A35. Although [{\feii}]~15586.550 is coincident in position with the stronger {\nitrogeni} line, it is far from the strongest [{\feii}] feature covered. The lack of detection in the ABE-A35 spectra of the stronger [{\feii}] lines rules those lines out as possibilities. Forbidden line emission in the optical was noted as early as 1976 for ABE-A35 \citep{1976A&A....47..293A}, but in the $H$-band the only clues suggesting the B[e] nature of this object are the abnormally strong {\hi}, {\feii}, and {\feii}-like emission lines.

\subsection{{\mgii} and {\siii}}
The lowest-energy {\mgii} and {\siii} lines covered in APOGEE data appear clearly in emission for ABE-A26 and are not confidently detected for any of the other stars. ABE-004, ABE-A05, and ABE-A29 possibly show exceedingly weak contributions from these lines, but blending renders the situation ambiguous in all cases aside from ABE-A26.

\begin{figure*}[htb!]
\includegraphics[width=18cm]{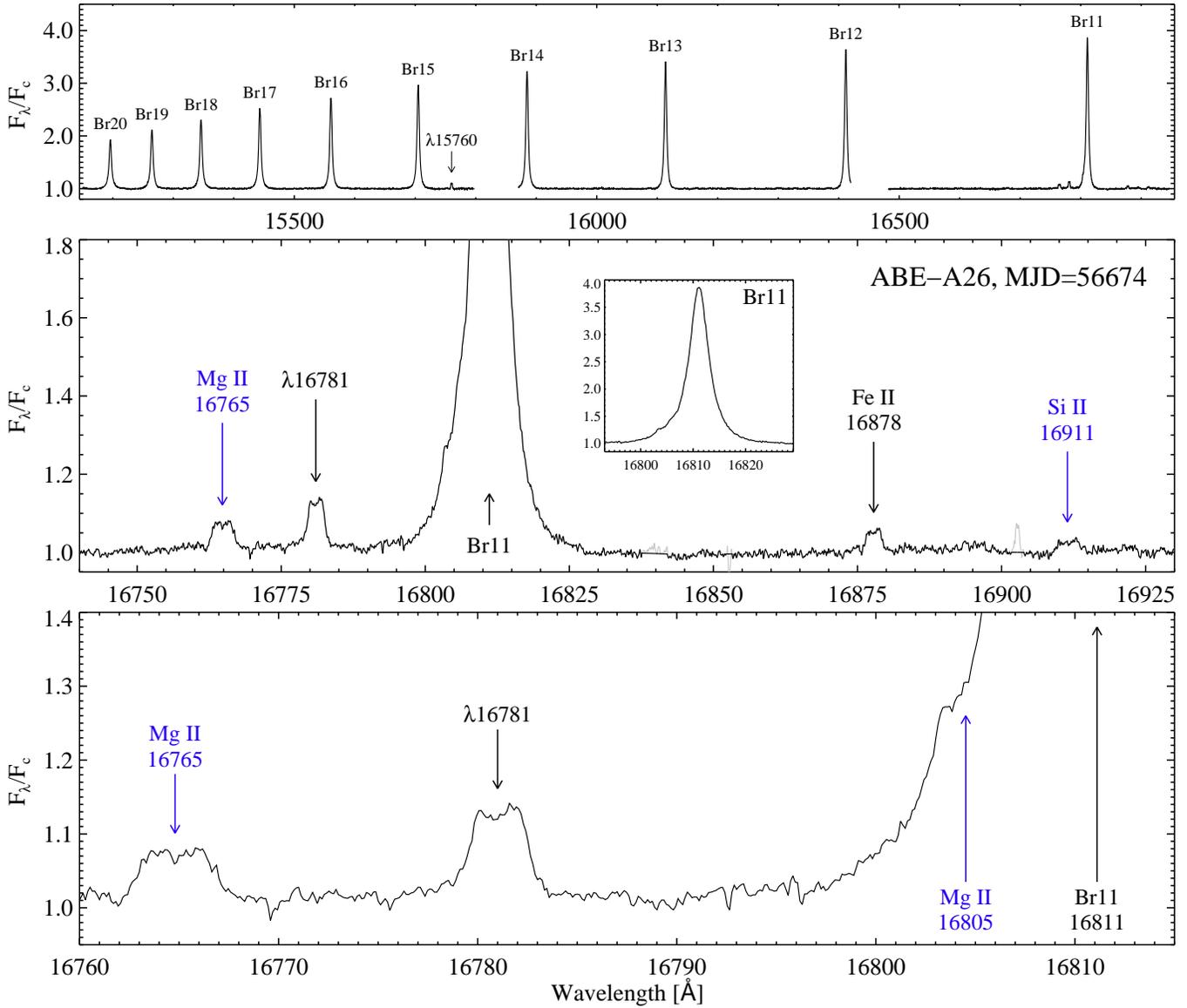}
\caption{{\mgii} and {\siii} emission lines in a spectrum of ABE-A26, the only star for which these lines are detected. While the full spectrum is presented in the upper panel, the lower two panels highlight the Br11 region and the weak metallic lines therein. The single-peaked Br11 line profile is displayed in the inset panel of the larger middle panel for comparison to the double-peaked profiles of the metallic lines. As expected from the detection of {\mgii}~16765,  the stronger line of this multiplet, {\mgii}~16804, appears blended with Br11 in the lowermost panel. A color version of this figure is available online. \label{fig_ionized}}
\end{figure*}

Figure~\ref{fig_ionized} displays a spectrum for ABE-A26 over differing wavelength ranges: the upper panel shows the full spectrum, while the middle and lower panels focus on the weak emission lines around Br11. Identification of the line blueward of $\lambda16781$ as {\mgii}~16764.80 requires that the stronger of three lines comprising this {\mgii} multiplet also be detected, and indeed the lower panel of Figure~\ref{fig_ionized} shows that {\mgii}~16804.52 is visible in the V wing of Br11 at apparently the correct intensity relative to {\mgii}~16764.80. Based on the intensities of these lines, the third {\mgii} line of the multiplet (16803.67~{\AA}) is not expected to appear and would overlap with {\mgii}~16804 anyway. The weak emission line redward of {\feii}~16878 in the middle panel of Figure~\ref{fig_ionized} is identified as Si~{\sc ii}~16911.430, the strongest Si~{\sc ii} line covered and a line with very similar energy levels to the {\mgii} lines (see Table~\ref{table_vpsummary}).

In addition to detection of the relatively high-ionization {\mgii} and {\siii} lines, the combination of single-peaked Brackett series lines and double-peaked metallic lines is unique to ABE-A26 among this sample. The double-peaked lines indicate that at least some of the circumstellar gas is organized in a disk. It is possible that ABE-A26 was observed by APOGEE during of after an outburst such that substantial Brackett series emitting gas is in the polar regions, leading to the single-peaked Brackett lines.

\section{Peak separations} \label{peaksep}

\subsection{Stars with abnormally large {\vp}} \label{magnetic}
Optical spectroscopy revealed that two stars, ABE-050 (HD~345439) and ABE-075 (HD~23478), with extremely large Brackett emission widths and double-peak separations are not classical Be stars \citep{2014arXiv1403.3239E}. Rather, these stars are analogues to the prototype magnetic emission B star $\sigma$ Orionis E first described as `Helium-rich' by \citet{1958ApJ...127..237G} and subsequently providing the first application \citep{2005ApJ...630L..81T} of the Rigidly Rotating Magnetosphere model of \citet{2005MNRAS.357..251T}. Large {\vp} for the Brackett series emission was a clue suggestive of a non-classical nature for these stars, but confirmation lay in the fact that both stars exhibit {\hi} emission well beyond the projected {\vsini} values (in these cases, a factor of two ore more beyond the projected {\vsini}). For classical Be stars with Keplerian disks, the velocity separations of emission peaks do not exceed 2{\vsini} \citep{1992A&AS...95..437D}.

\begin{figure*}[htb!]
\includegraphics[width=18cm]{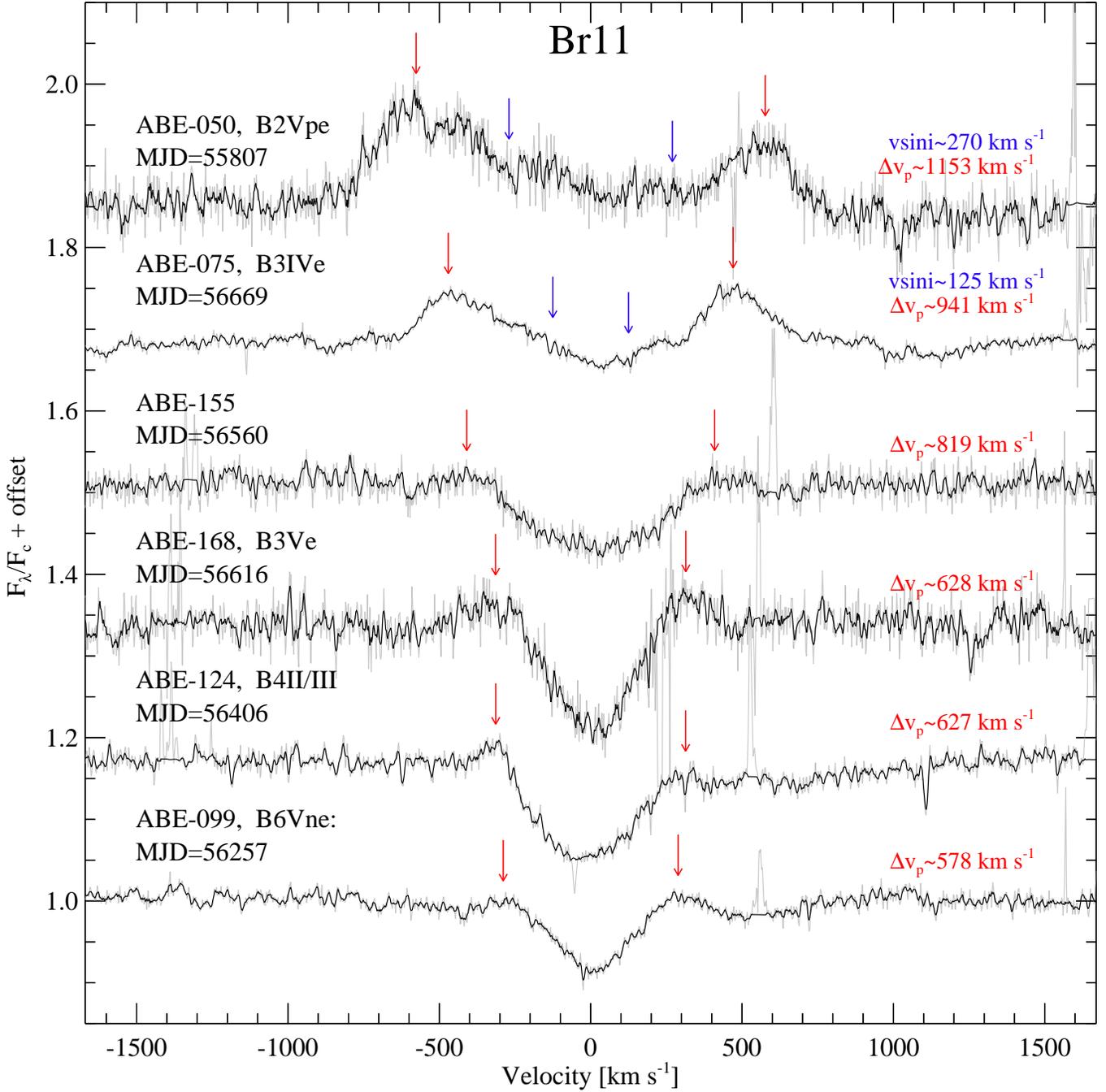}
\caption{Br11 line profiles for the ABE stars with the largest peak separations are shown. The {\vp} is listed and marked with arrows (red) for each source, while the {\vsini} measurements for ABE-050 and ABE-075 are given and indicated with arrows (blue) interior to the {\vp} arrows. Whereas ABE-050 and ABE-075 are confirmed $\sigma$ Orionis E type stars, the other four stars remain to be investigated further. A color version of this figure is available online. \label{fig_linew}}
\end{figure*}

Figure~\ref{fig_linew} compares the Br11 profiles of ABE-050 and ABE-075 to the stars with the next largest peak separations. Arrows indicate the measured peak separations and for ABE-050 and ABE-075, the inner sets of arrows indicate the {\vsini} values from \citet{2014arXiv1403.3239E}. As there are only a handful of magnetic B emission stars known to exist, it seems more likely that the ABE-155, ABE-168, ABE-124, and ABE-099 are weak-disked, edge-on classical Be stars rather than additional $\sigma$ Orionis E types. Either way, optical follow-up spectroscopy is required for proper diagnosis.

\subsection{Line-by-line {\vp} comparison} \label{vpeak}
\begin{figure*}[t!]
\includegraphics[width=18cm]{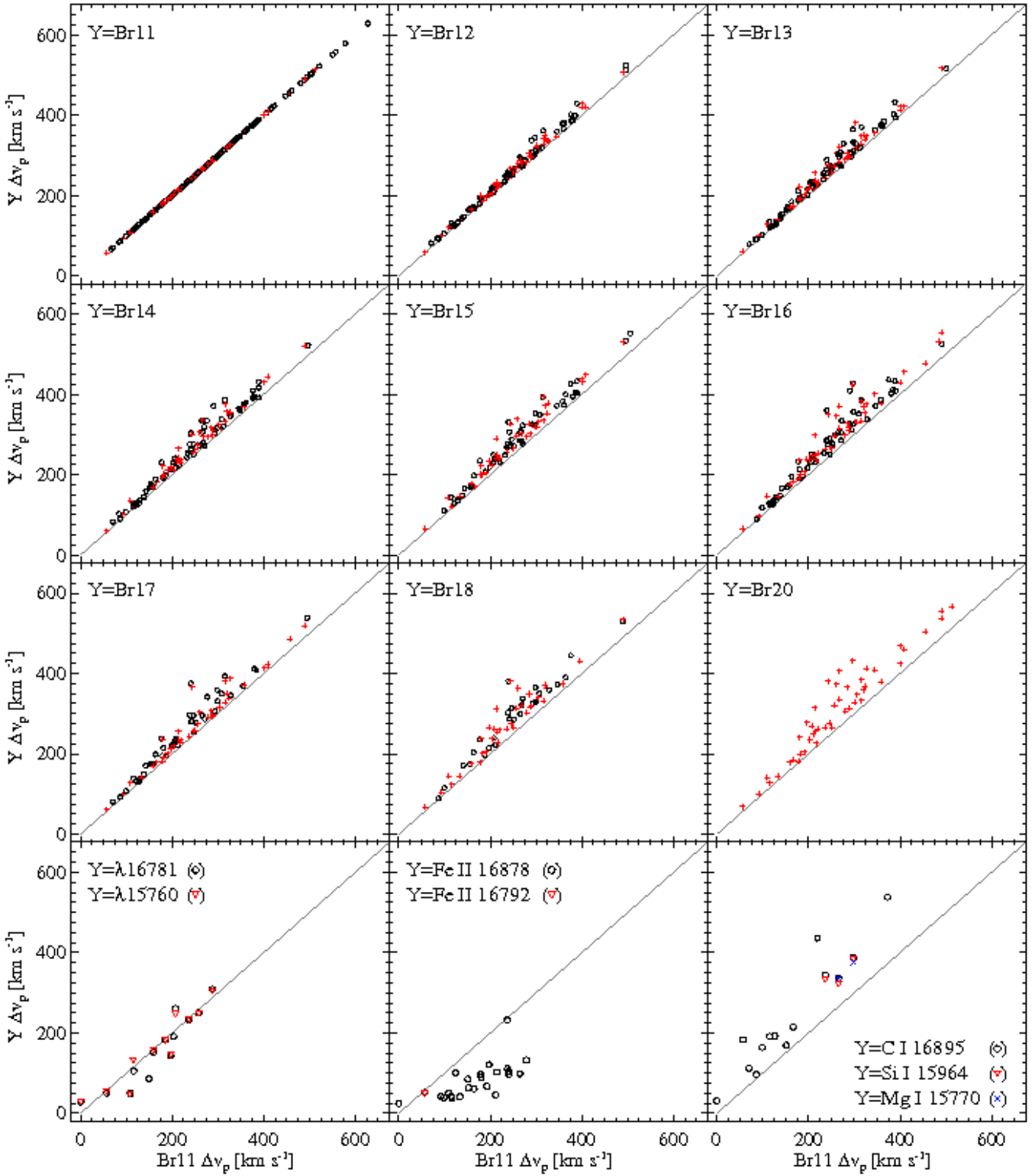}
\caption{The peak separation of Br11 is compared to the peak separations for the other Brackett series lines as well as the most routinely detected metallic emission lines. Symbol meanings are described in Section~\ref{vpeak}. A color version of this figure is available online. \label{fig_peaksep_stars}}
\end{figure*}

Figure~\ref{fig_peaksep_stars} plots the peak separations for Br11 versus the peak separations for Br12--Br18, Br20, $\lambda16781$, $\lambda15760$, {\feii}~16878, {\feii}~16792, {\ci}~16895, {\sii}~15964, and {\mgi}~15770. Each point represents the average peak separation for a line, from all spectra for a given star in which the Br11 {\vp} was measured in addition to the {\vp} of the line represented on the y-axis. In the upper nine panels, plus signs (red) correspond to stars for which the Br20 peak separation was measured and therefore to stars with strong or particularly sharp-peaked emission. The gaps between $\sim$400--500~{\kms} in the Br11 vs. Br18 and Br11 vs. Br17 panels are due to strong airglow lines impacting the emission peaks at large line width. High velocity gaps in the Br11 vs. Br12 and Br11 vs. Br14 panels are caused by either the Br12 V peak or the Br14 R peak falling too close to gaps between detectors. For Br13 and Br15, telluric absorption contamination is more likely for large line width. Grey lines indicate 1-to-1 relationships between the lines plotted in each panel.

The effect of increasing {\vp} toward weaker {\hi} lines is well-known for the Balmer \citep{1988A&A...190..187H} and Paschen \citep{1990A&AS...85..855A} lines, and the Brackett series lines are not an exception. Some stars (primarily the narrow-lined variety with Br11 $\Delta v_{\rm p} \textless 200${\kms}) show very little or no variation among Brackett series {\vp} but no convincing examples are found of decreasing {\vp} from Br11 toward Br20. We interpret the increasing peak separations toward weaker lines as kinematic in nature, such that the weaker Brackett lines (Br12--20) are simply formed closer to the rapidly-rotating central stars than e.g. Br11. The Br11--20 lines never take the form of winebottle-type profiles frequently observed in the optically-thick {\ha} line, where the effect of non-coherent scattering can produce inflections in the emission profile and effectively reduce the observed peak separation \citep{1992A&A...262L..17H}. Section 4 of \citet{1996A&AS..116..309H} provides a summary of the line broadening factors that contribute to Be star emission line profile shapes.

Based on the lower three panels of Figure~\ref{fig_peaksep_stars}, when the {\vp} of $\lambda15760$ and $\lambda16781$ are measured simultaneously, very similar values are found. The {\vp} for these lines are usually slightly smaller than the Br11 {\vp} but can be slightly larger as well. Excluding ABE-A26, for which Br11 is single-peaked, {\feii}~16878 is a strictly small-{\vp} line relative to Br11 with all {\vp} measurements less than 140~{\kms}. The {\ci}~16895 {\sii}~15964, and {\mgi}~15770 lines share nearly identical {\vp} in the available examples (ABE-084, ABE-188, ABE-A15; see Figure~\ref{fig_peculiar}), and all of the {\vp} measurements for {\ci}~16895 exceed the Br11 {\vp}.

\subsection{Line-emitting disk radii} \label{rdisk}

For Keplerian rotation in a gaseous disk, the orbital velocity decreases according to $r^{-1/2}$, where $r$ is the radial distance from star to disk. Given knowledge of the stellar rotational velocity, {\vsini}, the peak separation of an emission line can be used to calculate the approximate outer radius in the disk, {\rp}, at which that line is preferentially formed \citep{1969AcA....19..155S, 1972ApJ...171..549H, 1981AcA....31..395S, 1986MNRAS.218..761H}. Many authors \citep[e.g.,][]{1987A&A...173..299H, 1988A&A...190..187H, 1992A&AS...95..437D, 1990A&AS...85..855A, 1992ApJS...81..335S} have used this relation (Huang's law) to study the geometry of Be disks by estimating the individual line-emitting radii for {\hi} and metallic lines in the optical region. In units of stellar radii, {\rs}, {\rp} is calculated via Huang's Law \citep{1972ApJ...171..549H} as 

\begin{equation} \label{eq_rdisk}
r_{\rm d} = \left(\frac{2 \times v sin i}{\Delta v_{\rm p}}\right)^{2}
\end{equation}

where the equation is squared due to the assumption of a circular orbit. The resulting outer disk radii measurements for the subset of stars with available {\vsini} from the literature and Br11 or metallic line {\vp} measurements are listed in Table~\ref{table_rdisk}. In estimating {\rp}, the average {\vp} measured from all spectra for each star (the number of spectra used for each star is indicated in the `\# Obs' column) have been used. The relation in equation~\ref{eq_rdisk} may not necessarily hold for cases of emission lines with asymmetric peak intensities or for shell profiles so these instances have been noted in Table~\ref{table_rdisk}. In particular, large {\rp} estimates ($r_{\rm d} \textgreater 5 r_{\rm *}$) correspond to asymmetric and shell profiles. 

Taking the average of the Br11 {\rp} estimates for the 19 non-shell stars with roughly symmetric emission peaks, we find an average Br11 formation outer radius and associated standard deviation of

\begin{equation}
r_{\rm d}~({\rm Br}11) = 2.21{\pm}~0.73~R_{*}
\end{equation}

\citet{1988A&A...190..187H} found an average {\ha}-emitting radius of $\sim$20 {\rs}, while \citet{1992ApJS...81..335S} found an average of $\sim$19 {\rs}. It is important to note, however, that results from interferometry confirm that application of Huang's law to the double peaks of winebottle-type profiles (appearing for optically-thick lines like {\ha}) leads to artificially-large disc radii estimates \citep{1992A&A...262L..17H}. Interferometric studies typically produce radii estimates of less then 10 {\rs} over the optical and $JHK$ bands \citep[see disk radius measurements and papers referenced in Table 2 of][]{2013A&ARv..21...69}, more similar to what is found here from the Brackett lines.

For H$\gamma$ and {\feii} 6516~{\AA}, \citet{1992ApJS...81..335S} found emitting radii of $\sim$7.4 {\rs} and $\sim$3.9 {\rs} respectively. In a study of optical {\feii} emission lines for Be stars, \citet{2006A&A...460..821A} found that, on average, the optical {\feii} lines are formed at an outer disk radius of 2.0 stellar radii. Therefore, the Br11-emitting outer radius is roughly coincident with the optical {\feii}-emitting outer radius and well inside the {\ha}-emitting radius. 



Given the sparsity and potential wide-range of quality of {\vsini} information for our sample, disk radii are only estimated for the Br11 and metallic lines. However, it follows from Figure~\ref{fig_peaksep_stars} that the Br12--Br20 formation outer radii are interior to that of Br11. \citet{1990A&AS...85..855A} and \citet{1992ApJS...81..335S} found a correlation between formation location of individual optical lines and the upper energy levels ($E_{k}$) of the lines. Weaker lines with higher $E_{k}$ were generally found to have larger {\vp} and hence smaller {\rp}. This is also the case for the Brackett series lines, where $E_{k}$ increases slightly from Br11 to Br20. 

A trend toward large {\rp} is evident for the {\feii}~16878 line with respect to Br11. The average of the five available {\rp} estimates for {\feii} is $\sim$19~{\rs}, almost ten times the disk radius where Br11 is preferentially formed. A consequence of the widely-varying formation radii between Br11 and {\feii}~16878 is discussed in the following section.

\section{Single-epoch variation in V/R and RV} \label{vrratio}
As outlined in \citet{1991PASJ...43...75O}, long-term V/R variability for Be stars often entails shifts in radial velocity (RV) of entire emission line profiles toward whichever peak is stronger at the time and differences in V/R orientation between lines with different formation loci, such that V/R is necessarily constant from atomic species to atomic species or from line to line. These effects are believed to be caused by perturbations with the disks that give rise to one-armed global density waves that slowly precess through the disk with periods averaging 7 years \citep{2013A&ARv..21...69}.

Recent papers discussing the well known Be-shell star $\zeta$ Tau provided an example of V/R phase lags between Balmer and Brackett series lines and also between individual Brackett series lines. \citet{2007ApJ...656L..21W} hypothesized that the optical/NIR phase lag in V/R could be understood in terms of differing preferential formation radii and of the global density perturbation within the disk taking the form of a spiral arm. \citet{2009A&A...504..929S} and \citet{2009A&A...504..915C} subsequently showed this to be the case.

\subsection{{\hi} vs. {\hi} V/R phase lags} \label{hphaselag}
\begin{figure*}[htb!]
\includegraphics[width=18cm]{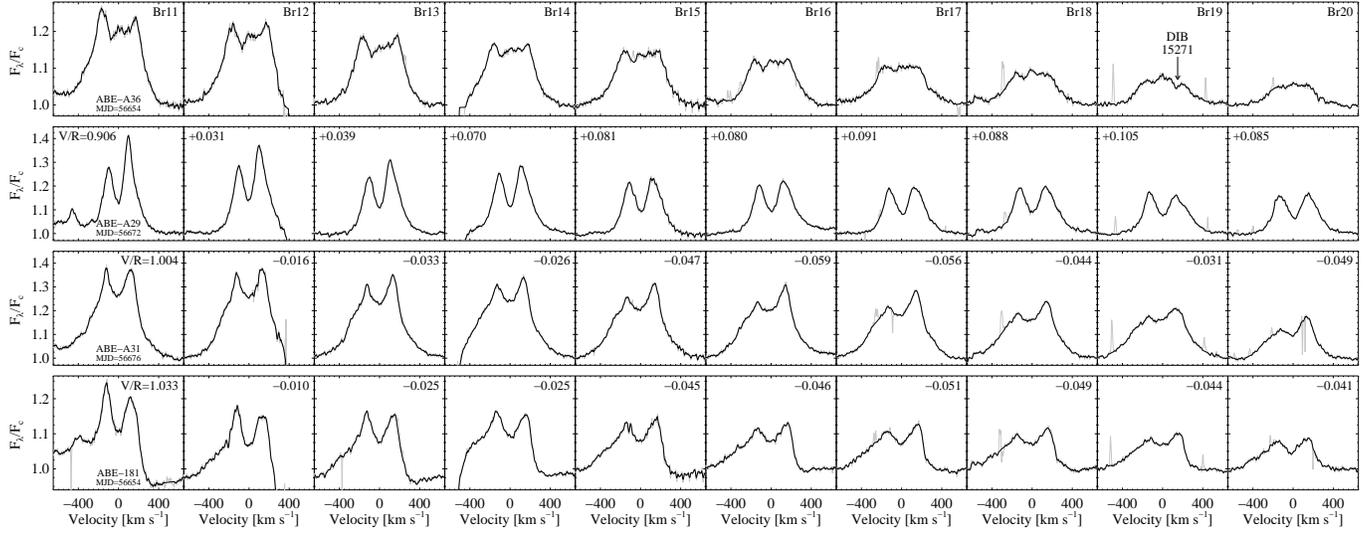}
\caption{Four examples of variation in V/R phase across the Brackett series lines. The Br11 profile for ABE-A36 has a quasi-triple-peaked morphology which gradually becomes a single-peaked morphology at Br20. For ABE-A29, ABE-A31, and ABE-181, the V/R ratio of Br11 is provided in the leftmost panels and subsequent panels provide the difference between the Br12--Br20 V/R ratios and the Br11 V/R ratio. Gradual changes in V/R orientation are seen among Brackett series lines for these stars: blue text for the differences means increasing V/R ratio from Br11 to Br20 (ABE-A29) while red text means decreasing V/R ratio from Br11 to Br20. DIB~15271 absorption is evident on the Br19 line for all four stars. \label{fig_weirdhlines}}
\end{figure*}

Evidence of V/R phase lags within the Brackett series lines is present in the spectra for the ABE stars represented in Figure~\ref{fig_weirdhlines}. Each Brackett line is displayed individually on a velocity scale in Figure~\ref{fig_weirdhlines} and, with the exception of ABE-A36 (discussed below), the V/R of Br11 for each spectrum is printed in left-most Br11 panels while the differences between the V/R of Br11 and the V/R of Br12--Br20 are printed in the Br12--Br20 panels. For ABE-A29, the V/R orientations progress from V{\textless}R at Br11 to V$\simeq$R at Br17. The Br19 profile is contaminated by DIB~15271 absorption, but the Br18 and Br20 profiles have V$\simeq$R similar to Br17. For ABE-A31 and ABE-181 the opposite progression takes place as the R peak increases in dominance from Br11 to Br20. Although the Br11 profiles for ABE-A36, ABE-A31, and ABE-181 are contaminated by underlying metallic emission ($\lambda16781$ and/or {\feii}~16878), the V/R phase lag is nonetheless plainly visible from comparison of the Br12 or Br13 profiles to Br20. 

Of the {\nstars} stars comprising the ABE sample, ABE-A36 is the only available example of quasi-triple-peaked (\textit{qtp}) Brackett series lines (see upper row of Figure~\ref{fig_weirdhlines}), where the ``quasi'' implies ambiguity as to whether there is actually a true third emission peak or whether instead the central absorption is split into multiple components. The V emission peak of Br11 for ABE-A36 is slightly higher than the R peak due to possible blending with $\lambda16781$, and the possible third peak appears between the dominant outer emission peaks with lesser intensity than those peaks. Br12 is similar in profile to Br11, but from Br12--Br17 the profiles gradually assume a `flat-topped' morphology with the apparent blue central absorption still weakly visible in contrast to the apparent red central absorption having all but disappeared. At Br18, evidence of the third (middle) emission peak emerges again, this time as the dominant peak since the outer V and R peaks have all but disappeared. Br19 is directly impacted by DIB~15271 but otherwise appears similar to Br18. Finally, the Br20 profile is smooth and rounded with only subtle traces of the blue central absorption and middle `emission peak' (the outer emission peaks are no longer visible).

\citet{2009A&A...504..929S} pointed out that although \textit{qtp} H$\alpha$ profiles occur at certain times during the V/R cycle of $\zeta$ Tau, optically thin lines likes O~{\sc i}~8446 and the Brackett series never exhibited any evidence of \textit{qtp}. It is therefore unusual that \textit{qtp} profiles are observed in the Brackett lines for ABE-A36. We can report no additional examples.

\subsection{{\hi} vs. metallic V/R and RV phase lags} \label{vrmismatch}
\begin{figure*}[htb!]
\includegraphics[width=18cm]{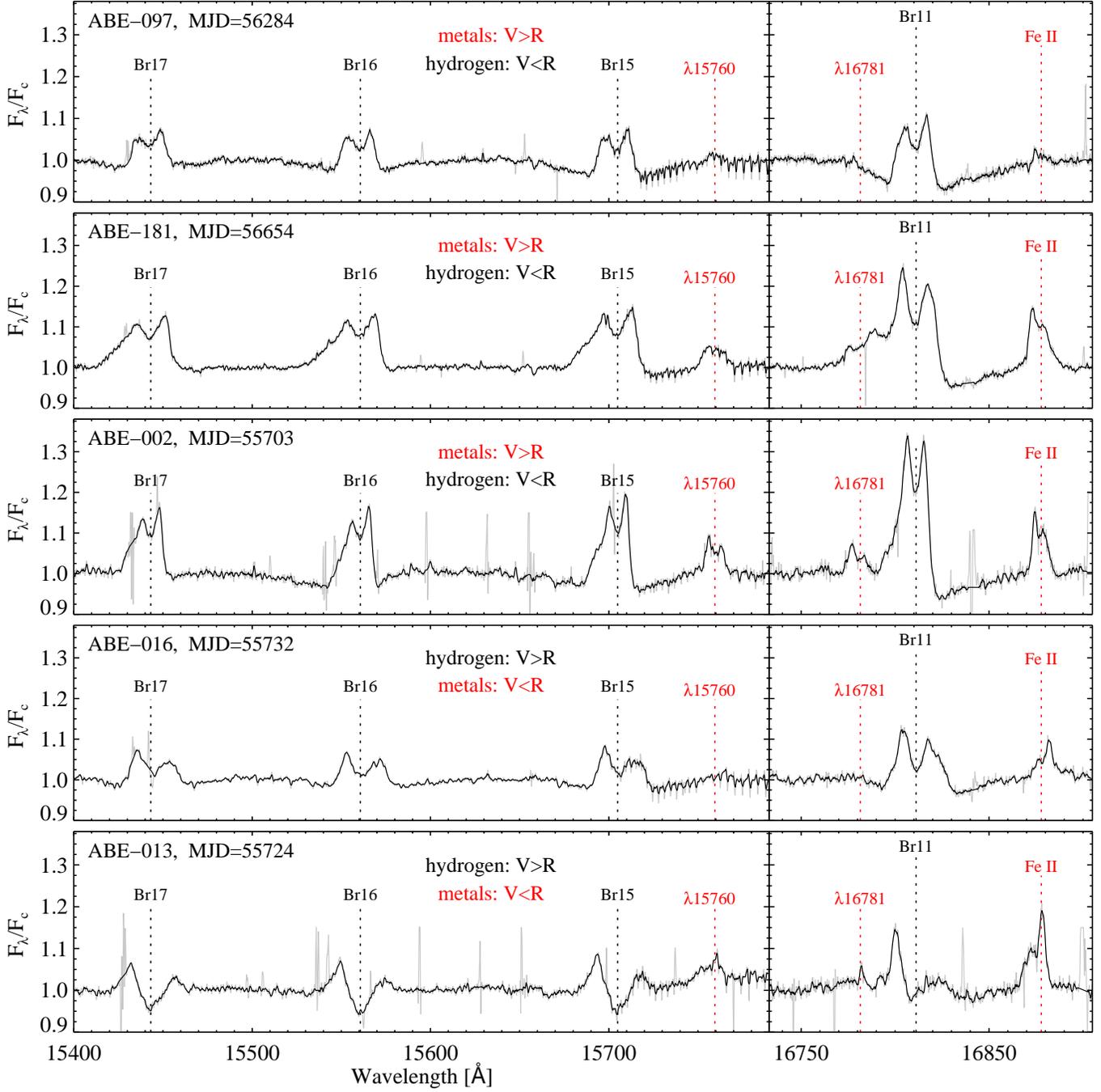}
\caption{{\hi} versus metallic V/R orientation mismatches are evident in the APOGEE spectra of ABE-097, ABE-181, ABE-002, ABE-016, and ABE-013. The emission wings for ABE-181 and ABE-002 are also clearly extended in the direction of the weaker emission peak for each line, and the metallic emission profiles for ABE-013 appear to be offset in radial velocity from the Brackett lines. A color version of this figure is available online. \label{fig_vrmismatch}}
\end{figure*}

The five examples shown in Figure~\ref{fig_vrmismatch} represent the first known examples of disagreement between the V/R orientations of Brackett series versus metallic lines ({\feii}~16878, $\lambda15760$, and $\lambda16781$). ABE-097, ABE-181, and ABE-002 have V$\textless$R for {\hi} and V$\textgreater$R for metallic lines, while ABE-016 and ABE-013 have V$\textgreater$R for {\hi} and V$\textless$R for metallic lines. Due to the contaminated Br11 profiles for ABE-013, ABE-002, and ABE-181, where the V peak height has been increased by underlying {\feii}~16792 emission, the left-hand panels of Figure~\ref{fig_vrmismatch} are extended to encompass not only $\lambda15760$, but also Br15--Br17 to show the typical {\hi} V/R orientation for each star.

Although the lack of available stellar absorption lines means that precise stellar RV determination is not possible, the spectrum for ABE-013 in Figure~\ref{fig_vrmismatch} has been corrected to rest frame based on the average positions of the deep absorptions in the Br12--Br20 lines, while the other spectra lack the deep {\hi} absorptions and therefore were corrected for Doppler shift based on average emission peak shift for the Brackett lines. The result that emerges for ABE-013 is that the Brackett series absorptions do not coincide in RV not with the central depressions in the metallic emission lines as is normally true (see Figure~\ref{fig_unidentified}), but instead the Brackett series absorptions coincide in RV with the R emission peaks of the metallic lines. More specifically, the {\feii}~16878, $\lambda15760$, and $\lambda16781$ profiles are shifted in RV with respect to {\hi} absorption by $\sim$50~{\kms} in the direction of stronger {\hi} peaks as expected from \citet{1991PASJ...43...75O}. 

ABE-002 and ABE-181 exhibit clear evidence of V/R-related RV shifts in the emission profiles, though of a slightly different variety from that of ABE-013. Metallic and {\hi} emission wings for both stars are conspicuously enhanced on the side of the line profiles opposite the stronger emission peak for ABE-002 and ABE-181, with the {\hi} wings being enhanced on the blue side and the metallic wings enhanced on the red side. The enhanced blue wings suggest that significantly more emission is being formed in the inner regions of the approaching side of the disk, and the steep declines in intensity, from stronger emission peak to narrower emission base (R side of {\hi} for ABE-002 and ABE-181), imply cavities in the inner regions of the receding sides of the disks and relatively increased emission coming from the outer regions of the disks. We interpret these line profiles to suggest more tightly wound spiral patterns to the density oscillation in the disks of these stars versus $\zeta$ Tau.

\section{Br11 line profiles} \label{lineprofs}
\begin{figure*}
\includegraphics[width=18cm]{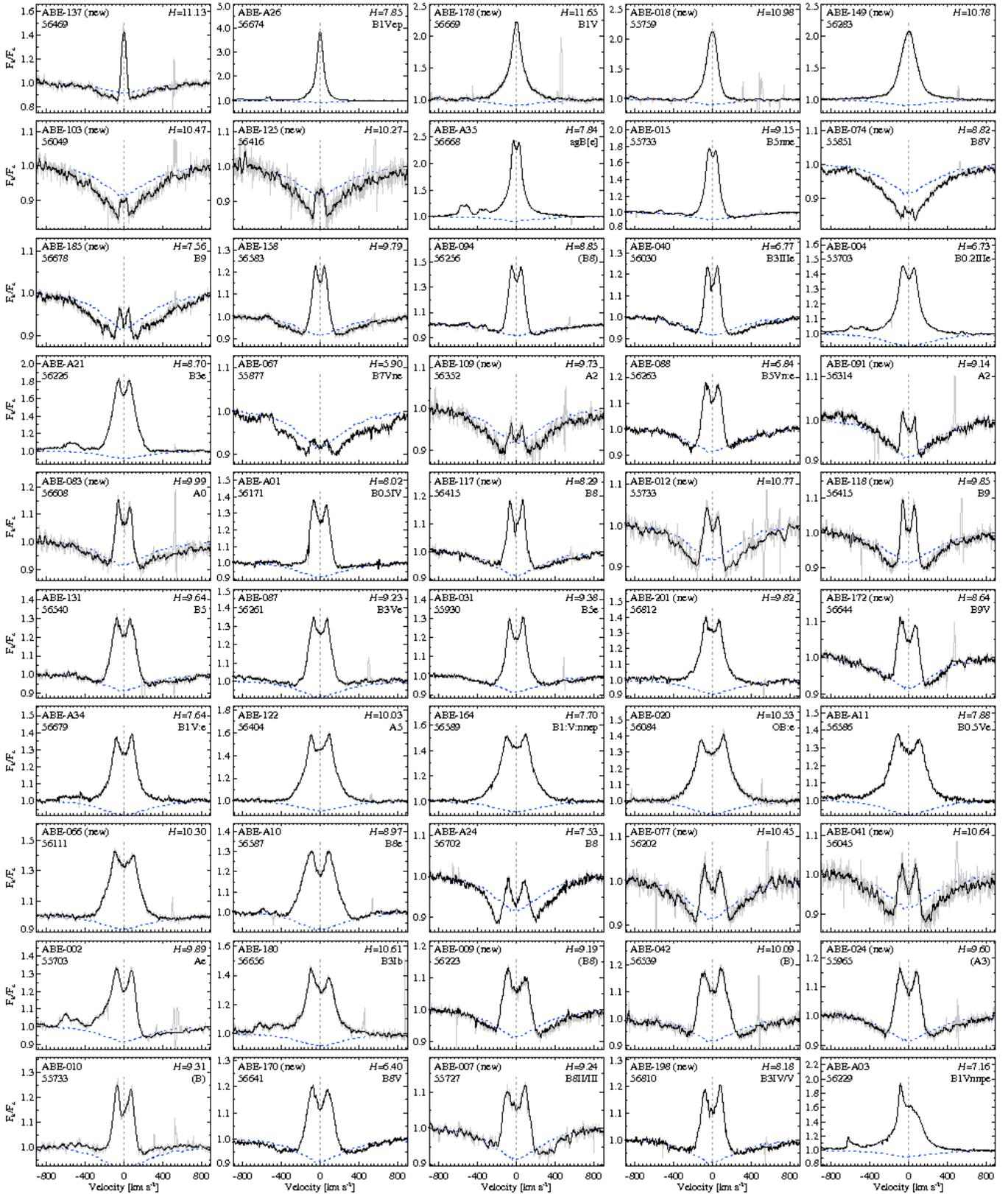}
\caption{Br11 line profiles sorted approximately according to inclination angle, from pole-on to edge-on. ABE identifiers, observation MJDs, 2MASS $H$ magnitudes, and literature spectral types (where available) are printed in each panel. The average Br11 profile of the quiescent Be stars (ABE-Q01--ABE-Q23) is displayed as a dotted line (blue), and vertical dotted lines (grey) indicate emission peak midpoints or estimated line centers if emission peaks are not present. Color versions of this figure as well as Figures~\ref{fig_isort_pt2}--\ref{fig_isortC} are available online. \label{fig_isort_pt1}}
\end{figure*}

\begin{figure*}
\includegraphics[width=18cm]{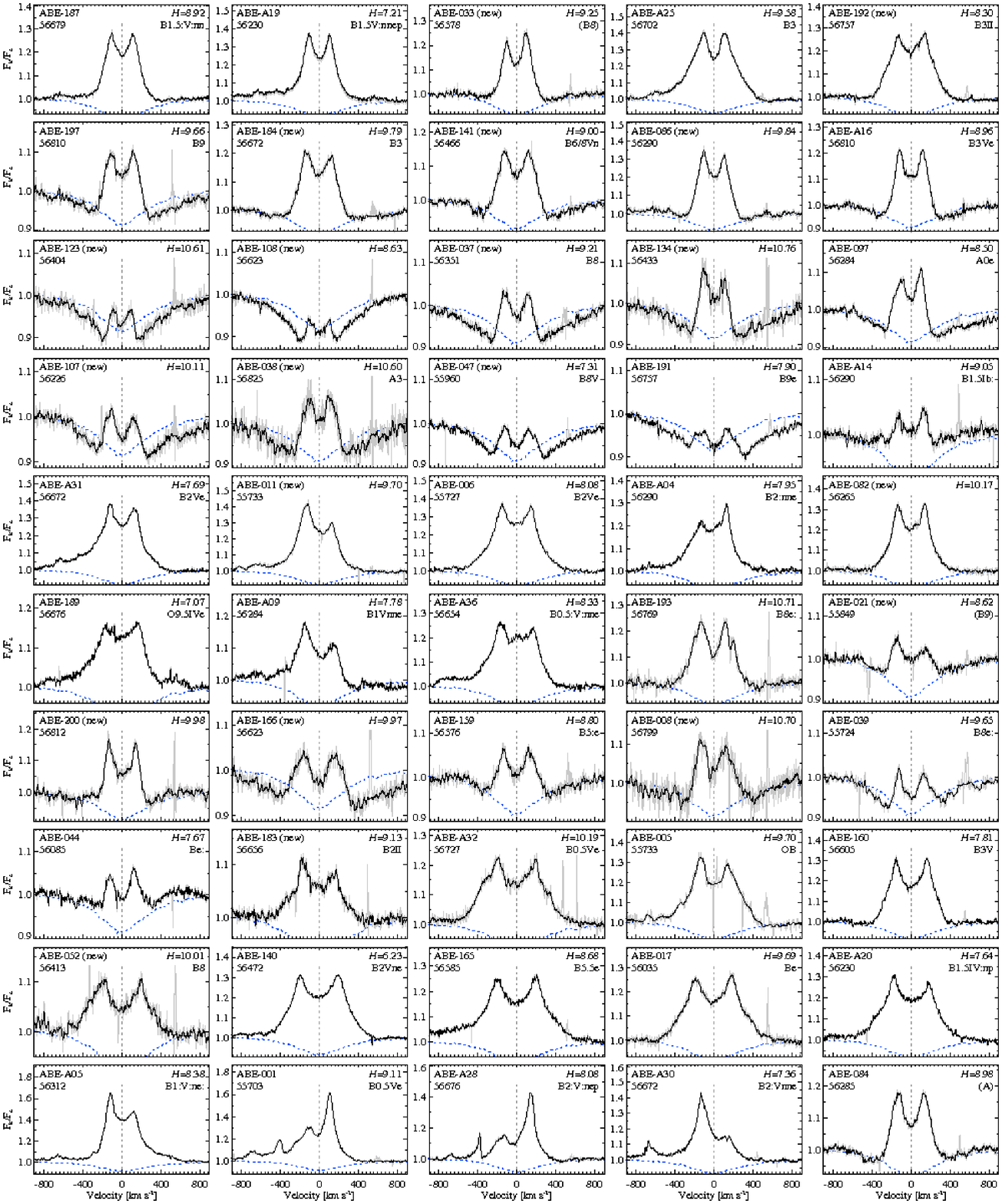}
\caption{Br11 line profiles sorted by inclination angle, going from pole-on to edge-on. Meanings are the same as in Figure~\ref{fig_isort_pt1}. \label{fig_isort_pt2}}
\end{figure*}

\begin{figure*}
\includegraphics[width=18cm]{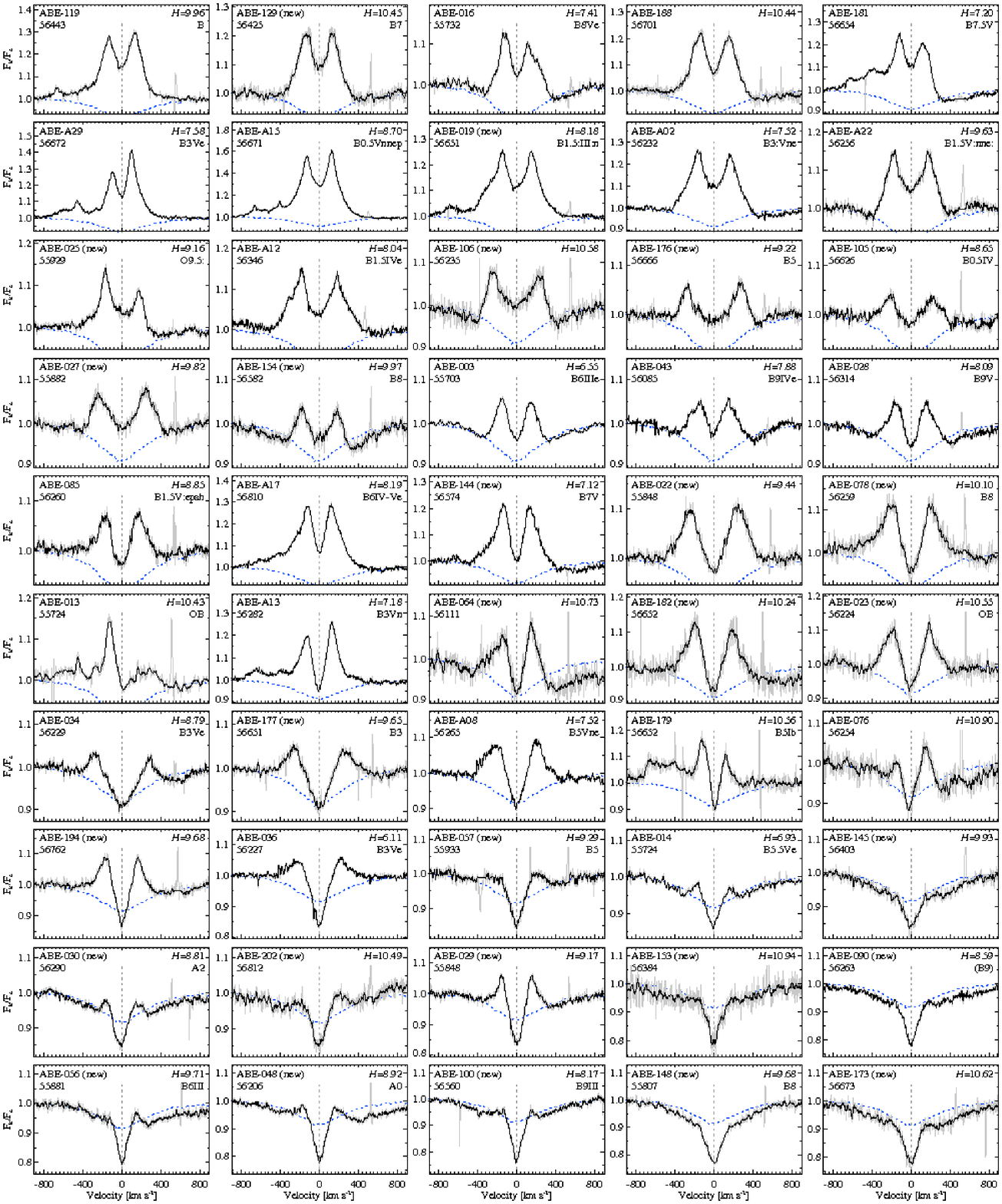}
\caption{Br11 line profiles sorted by inclination angle, going from pole-on to edge-on. Meanings are the same as in Figure~\ref{fig_isort_pt1}. \label{fig_isort_pt3}}
\end{figure*}

\begin{figure*}
\includegraphics[width=18cm]{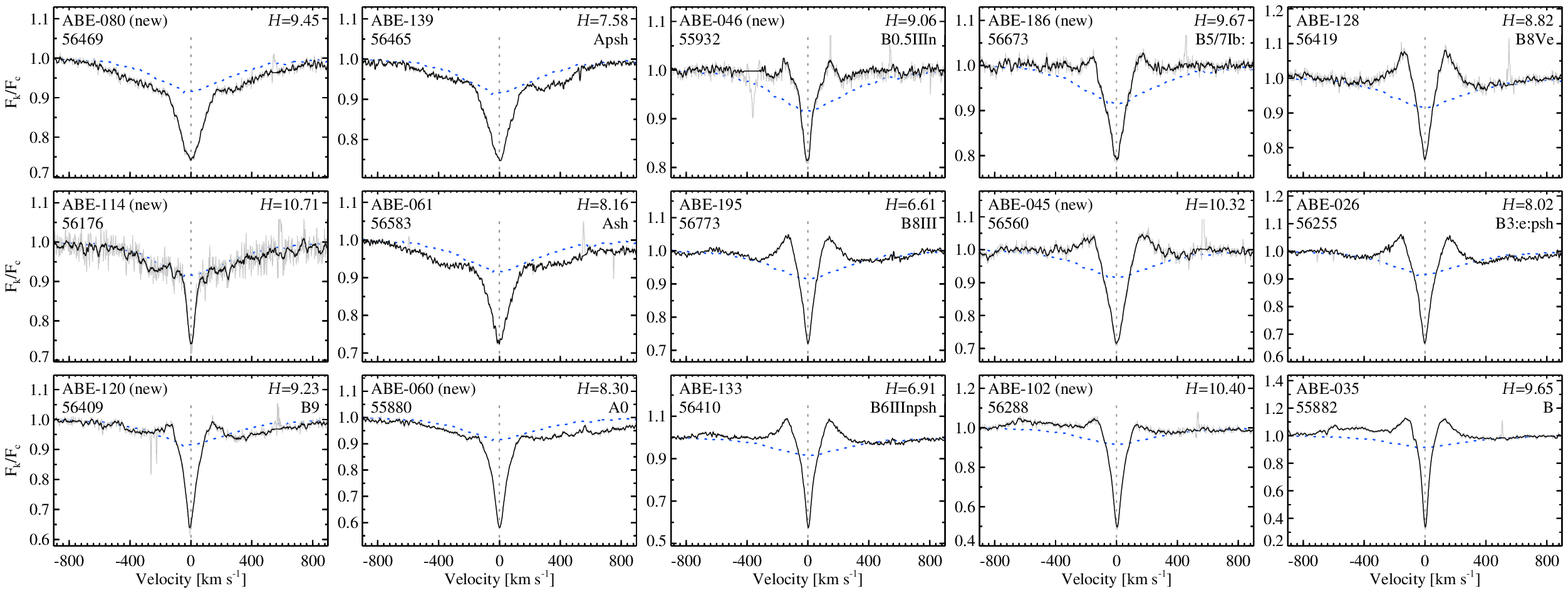}
\caption{Br11 line profiles sorted by inclination angle, going from pole-on to edge-on. Meanings are the same as in Figure~\ref{fig_isort_pt1}. \label{fig_isort_pt4}}
\end{figure*}

\begin{figure*}
\includegraphics[width=18cm]{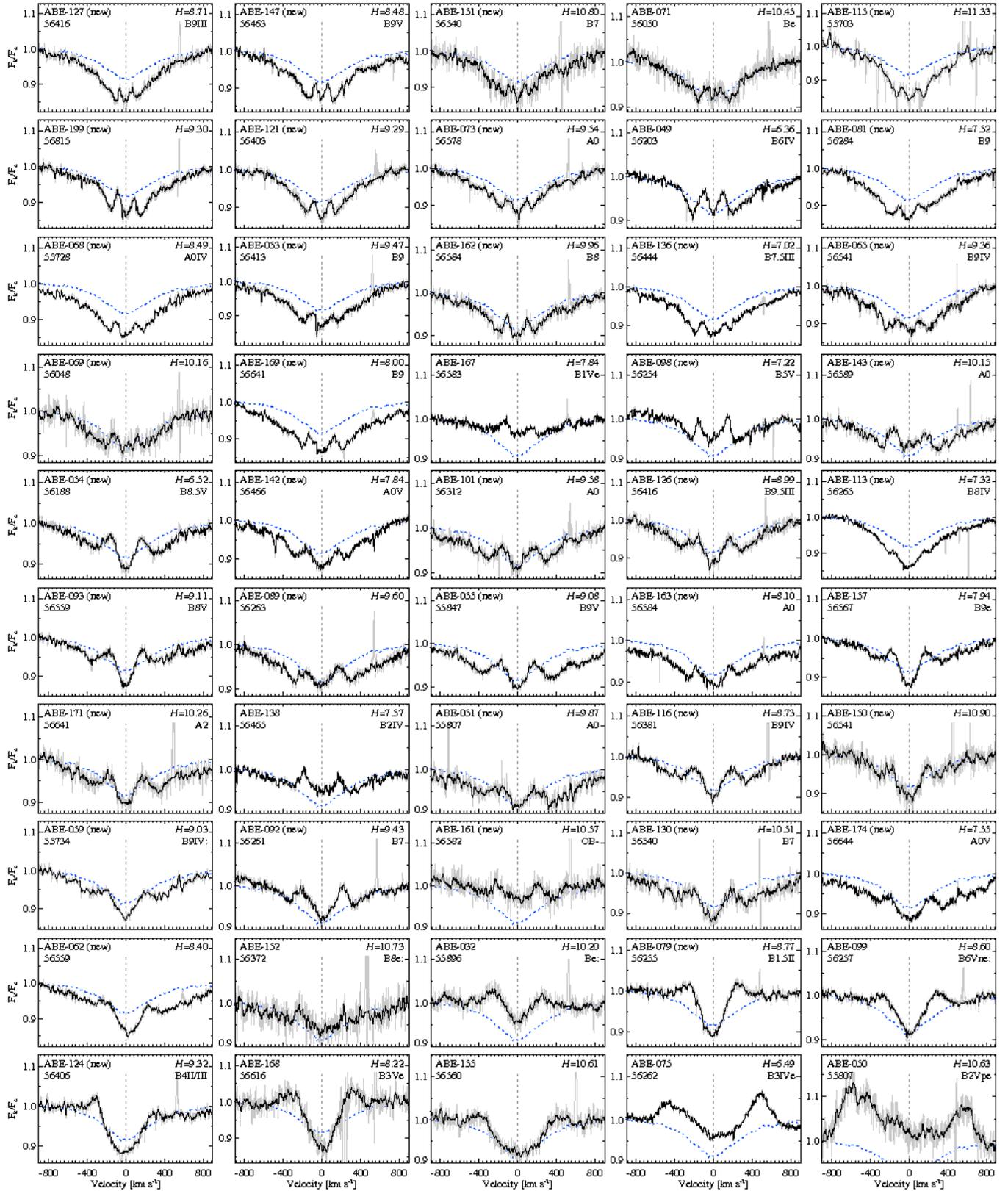}
\caption{Br11 line profiles for stars with weak or ambiguous emission profile type, as well as for the $\sigma$ Ori E type stars ABE-075 and ABE-050. The panels are sorted by Br11 peak separation. Meanings are otherwise the same as in Figure~\ref{fig_isort_pt1}. \label{fig_isortB}}
\end{figure*}

\begin{figure*}
\includegraphics[width=18cm]{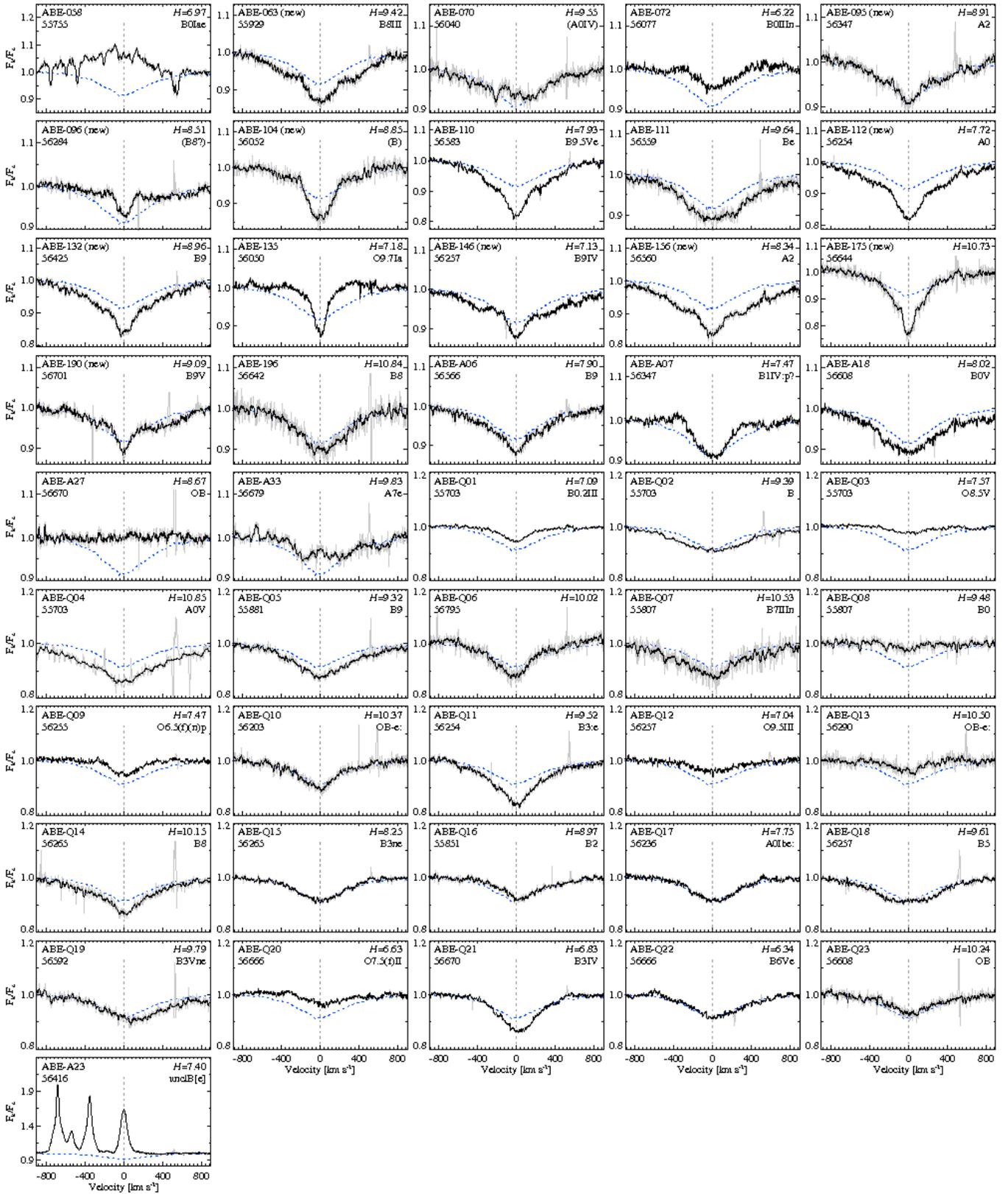}
\caption{Br11 line profiles for stars with weak or ambiguous emission profile type and lack of discernible emission peaks, followed by the sample of previously-known emission stars that produced little or no emission in the APOGEE spectra. The panels are mostly sorted by ABE identifier (followed by ABE-A23). Note that the telluric correction is problematic for the plug-plate on which ABE-058 was observed; Br11 is clearly in emission, but the profile is badly contaminated by telluric absorption features. Meanings are otherwise the same as in Figure~\ref{fig_isort_pt1}. \label{fig_isortC}}
\end{figure*}

Br11 line profiles from the highest-quality-available spectrum of each ABE star are displayed in Figures~\ref{fig_isort_pt1}--\ref{fig_isortC}. In Figures~\ref{fig_isort_pt1}--\ref{fig_isort_pt4}, the Br11 profiles of 165 ABE stars are qualitatively sorted by profile type, going from single-peaked and narrow double-peaked profiles to deep shell profiles. According to the models of \citet{1992A&A...262L..17H} and \citet{2000A&A...359.1075H}, the major line profile shape differences for Be stars are an effect of the inclination angles ($i$) at which the circumstellar disks are observed. \citet{1996A&AS..116..309H} used high-resolution {\ha} and optical {\feii} profiles to devise a Be sub-classification scheme based on the notion of $i$ dictating to a large extent line profile morphology. \citet{2010ApJS..187..228S} later showed that line shape in an optically-thick line like {\ha} is not dictated solely by $i$ and that very different profile shapes may be observed at fixed $i$, but no such investigations of the Brackett series lines have been done. 

In sorting the Br11 profiles of the ABE stars according to expected $i$, we relied largely on the models of optically thin lines from \citet{1992A&A...262L..17H}. The most readily classified Br11 profiles correspond to $i\sim0^{\circ}$ (pole-on), where single-peaked or narrow double-peaked emission is expected, and $i\sim90^{\circ}$ (edge-on), where deep shell absorption with a sharp core (with or without adjacent emission) is expected. The situation is far more ambiguous for profiles corresponding to intermediate $i$, but a general trend of increasing central depression depth and overall line width with increasing $i$ is apparent. Line profiles that could not be satisfactorily sorted by $i$, due to weakness or ambiguity of the disk features, are shown in Figures~\ref{fig_isortB} and \ref{fig_isortC}. Figure~\ref{fig_isortB} profiles are sorted by Br11 peak separation, and Figure~\ref{fig_isortC} profiles are sorted by ABE identifier.

\section{Conclusions} \label{conclusions}
SDSS-III/APOGEE has serendipitously provided the first high-resolution view of the $H$-band properties of a large number of Be stars, the majority of which are targeted quasi-randomly by the survey as telluric calibrators. Although significant progress has been made toward understanding Be stars over the past few decades via high-resolution optical, interferometric, and spectropolarimetric studies \citep{2013A&ARv..21...69}, any fully explanatory model of the classical Be phenomenon will need to account for the multi-wavelength properties of these stars. Multi-wavelength studies of statistically-significant samples of Be stars are critical yet have historically been few and far between, though the limited exceptions \citep{2000A&AS..141...65C, 2001A&A...371..643S} have been highly valuable. Due to simultaneous coverage in the $H$-band of numerous {\hi} lines that are minimally affected by underlying photospheric absorption in comparison to the Balmer series lines, the $H$-band is particularly promising in terms of utility toward V/R variability and general Be disk studies. Despite the $H$-band covering only a limited number of metallic emission lines, we have shown that the {\feii} and {\feii}-like ($\lambda15760$ and $\lambda16781$) lines are highly interesting in the context of V/R variability and phase lags between various atomic species. 

In the first of a series of papers exploring the $H$-band properties of Be stars, we have identified the non-hydrogen emission line content of the APOGEE Be star spectra, analyzed the kinematic properties of the metallic and {\hi} features, and discussed the more exceptional Be stars within the sample as well as those deviating from the typical emission line content. Further investigation of the identities of emission lines at 15760~{\AA} and 16781~{\AA} is needed, but may require updated atomic line lists. Since little is known about most of the ABE stars themselves, including spectral type and rotation speed, optical follow-up study of these stars is also needed in order to develop a better understanding of $H$-band properties as they relate to known stellar parameters.

\vspace{0.5cm}
\scriptsize{\emph{Acknowledgements.}} Funding for SDSS-III has been provided by the Alfred P. Sloan Foundation, the Participating Institutions, the National Science Foundation, and the U.S. Department of Energy Office of Science. The SDSS-III web site is http://www.sdss3.org/.

SDSS-III is managed by the Astrophysical Research Consortium for the Participating Institutions of the SDSS-III Collaboration including the University of Arizona, the Brazilian Participation Group, Brookhaven National Laboratory, Carnegie Mellon University, University of Florida, the French Participation Group, the German Participation Group, Harvard University, the Instituto de Astrofisica de Canarias, the Michigan State/Notre Dame/JINA Participation Group, Johns Hopkins University, Lawrence Berkeley National Laboratory, Max Planck Institute for Astrophysics, Max Planck Institute for Extraterrestrial Physics, New Mexico State University, New York University, Ohio State University, Pennsylvania State University, University of Portsmouth, Princeton University, the Spanish Participation Group, University of Tokyo, University of Utah, Vanderbilt University, University of Virginia, University of Washington, and Yale University. 

We thank the anonymous referee, and Kevin Covey, both of whom provided feedback which substantially improved the paper. The first author additionally thanks his mother for proofreading drafts of the paper.


\newpage
\clearpage
\begin{longtable}{cccccc}
\caption{\label{table_stars} List of ABE stars}\\
\hline\hline
ABe ID & Star Name & 2MASS & Lit. Spec. Type & Ref. & {\vp}Br11 \\
 & & mag & & & km s$^{-1}$ \\
\hline
\endfirsthead
\caption{continued.}\\
\hline\hline
ABe ID & Star Name & 2MASS & Lit. Spec. Type & Ref. & {\vp}Br11 \\
 & & mag & & & km s$^{-1}$ \\
\hline
\endhead
\hline
\endfoot
\tiny{001} & \tiny{VES 203} & \tiny{9.108} & \tiny{B0.5Ve} & \tiny{74} & \tiny{208} \\
\tiny{002} & \tiny{HD~228576} & \tiny{9.888} & \tiny{Ae} & \tiny{29} & \tiny{152} \\
\tiny{003} & \tiny{HR 7757} & \tiny{6.548} & \tiny{B6IIIe} & \tiny{22} & \tiny{305} \\
\tiny{004} & \tiny{HD~229221} & \tiny{6.734} & \tiny{B0.2IIIe} & \tiny{66} & \tiny{115} \\
\tiny{005} & \tiny{Hen 3-1876} & \tiny{9.699} & \tiny{OB} & \tiny{6} & \tiny{271} \\
\tiny{006} & \tiny{HD~179405} & \tiny{8.084} & \tiny{B2Ve} & \tiny{68} & \tiny{294} \\
\tiny{\textbf{007}} & \tiny{HD~178920} & \tiny{9.236} & \tiny{B8II/III} & \tiny{58} & \tiny{181} \\
\tiny{\textbf{008}} & \tiny{J18000176-2323071} & \tiny{10.699} & \tiny{\nodata} & \tiny{\nodata} & \tiny{233} \\
\end{longtable}
\tiny{[1] \citet{1942ApJ....96...15M}; [2] \citet{1949AnHar.112....1C}; [3] \citet{1949ApJ...110..387M}; [4] \citet{1950ApJ...111..495P}; [5] \citet{1951ApJ...113..624M}; [6] \citet{1952ApJ...115..459N};} \\
\tiny{[7] \citet{1953ApJ...118..318M}; [8] \citet{1955ApJS....2...41M}; [9] \citet{1956AN....283..109H}; [10] \citet{1956ApJS....2..389H}; [11] \citet{1958JO.....41...43D}; [12] \citet{1958TrRig...7...33A};} \\ \tiny{
[13] \citet{1959ApJS....4....1M}; [14] \citet{1959LS....C01....0H}; [15] \citet{1961AnTou..28...33B}; [16] \citet{1962JO.....45..349F}; [17] \citet{1963ArA.....3...97R}; [18] \citet{1963MmRAS..68..173F};} \\ \tiny{
[19] \citet{1967AJ.....72.1199M}; [20] \citet{1967PASP...79..181S}; [21] \citet{1968AJ.....73..233R}; [22] \citet{1968ApJS...17..371L}; [23] \citet{1968PASP...80..197G}; [24] \citet{1970MmRAS..73..153W};} \\ \tiny{
[25] \citet{1971ApJS...23..257W}; [26] \citet{1972AJ.....77..750C}; [27] \citet{1973A&A....22..229L}; [28] \citet{1976ApJ...210...65T}; [29] \citet{1976ApJS...30..491H}; [30] \citet{1976KiIND.........V};} \\ \tiny{
[31] \citet{1977ApJ...213..105D}; [32] \citet{1977ApJ...217..127C}; [33] \citet{1977ApJS...33..459S}; [34] \citet{1977MNRAS.180..691H}; [35] \citet{1977PW&SO...2...71S}; [36] \citet{1978AJ.....83..172R};} \\ \tiny{
[37] \citet{1979AbaOB..51....1B}; [38] \citet{1979RA......9..479C}; [39] \citet{1980BICDS..19...74O}; [40] \citet{1982IAUS...98..261J}; [41] \citet{1982mcts.book.....H}; [42] \citet{1985cbvm.book.....V};} \\ \tiny{
[43] \citet{1988AJ.....95.1543B}; [44] \citet{1988PASP..100.1084B}; [45] \citet{1988mcts.book.....H}; [46] \citet{1989AN....310..223R}; [47] \citet{1990A&AS...85.1069S}; [48] \citet{1992AJ....104.1132T};} \\ \tiny{
[49] \citet{1992ApJS...81..795G}; [50] \citet{1993A&AS...97..755T}; [51] \citet{1994AJ....107.1556G}; [52] \citet{1995A&AS..110..367N}; [53] \citet{1995ApJS...99..135A}; [54] \citet{1997AAHam..11.....K};} \\ \tiny{
[55] \citet{1998A&A...340..117L}; [56] \citet{1998MNRAS.298..185E}; [57] \citet{1999A&AS..137..451G}; [58] \citet{1999mctd.book.....H}; [59] \citet{2001A&A...368..912Y}; [60] \citet{2001A&A...378..861C};} \\ \tiny{
[61] \citet{2001KFNT...17..409K}; [62] \citet{2002A&A...386..709F}; [63] \citet{2002ApJ...573..359A}; [64] \citet{2003A&A...412..219M}; [65] \citet{2004AJ....127.1682H}; [66] \citet{2004AN....325..380N};} \\ \tiny{
[67] \citet{2004AN....325..749N}; [68] \citet{2006A&A...451.1053F}; [69] \citet{2006MNRAS.371..252L}; [70] \citet{2007ApJ...658.1264U}; [71] \citet{2010A&A...522A.107R}; [72] \citet{2010AJ....139.1283W};} \\ \tiny{
[73] \citet{2011ApJS..193...24S}; [74] \citet{2011BASI...39..517M}; [75] \citet{2012A&A...541A..34S}; [76] \citet{2013yCat.3271....0C}; [77] \citet{2014arXiv1403.3239E}}
\end{longtable}

\newpage
\clearpage
\begin{deluxetable}{lccrrrrccccc}
\tablecolumns{12}
\setlength{\tabcolsep}{0.02in}
\tablewidth{0pc}
\tablecaption{Observed Emission Lines and Summary of {\vp} Measurements \label{table_vpsummary}}
\tablehead{\colhead{(1)}               & \colhead{(2)}                              & \colhead{(3)}                               & \colhead{(4)}                    & \colhead{(5)}                  & \colhead{(6)}                  & \colhead{(7)}                    & \colhead{(8)}                     & \colhead{(9)}                & \colhead{(10)}                  & \colhead{(11)}                  & \colhead{(12)}                \\
           \colhead{\scriptsize{Atom}} & \colhead{\scriptsize{$\lambda_{\rm vac}$}} & \colhead{\scriptsize{$\lambda_{\rm vac}$}}  & \colhead{\scriptsize{Diff.}}     & \colhead{}                     & \colhead{}                     &                                  & \colhead{\scriptsize{N}}          & \colhead{\scriptsize{{\vp}}} & \colhead{\scriptsize{{\vp}}}    & \colhead{\scriptsize{{\vp}}}    & \colhead{\tiny{Other}}        \\
           \colhead{\scriptsize{or}}   & \colhead{\scriptsize{lab}}                 & \colhead{\scriptsize{obs\tablenotemark{a}}} & \colhead{\scriptsize{lab--obs}}  & \colhead{\scriptsize{$E_{i}$}} & \colhead{\scriptsize{$E_{k}$}} &                                  & \colhead{\scriptsize{detections}} & \colhead{\scriptsize{N}}     & \colhead{\scriptsize{range}}    & \colhead{\scriptsize{mean}}     & \colhead{\tiny{possible}}     \\
           \colhead{\scriptsize{ion}}  & \colhead{\scriptsize{[{\AA}]}}             & \colhead{\scriptsize{[{\AA}]}}              & \colhead{\scriptsize{[{\AA}]}}   & \colhead{\scriptsize{[eV]}}    & \colhead{\scriptsize{[eV]}}    & \colhead{\scriptsize{log($gf$)}} & \colhead{\tiny{yes (maybe)}}      & \colhead{\scriptsize{stars}} & \colhead{\scriptsize{[{\kms}]}} & \colhead{\scriptsize{[{\kms}]}} & \colhead{\tiny{contribution}}  }
\startdata
\scriptsize{{\hi} (Br20)} & \scriptsize{15195.996} & \scriptsize{15195.932} & \scriptsize{$0.064$} & \scriptsize{12.749} & \scriptsize{13.564} & \scriptsize{$-1.487$} & \scriptsize{\nodata} & \scriptsize{48}  & \scriptsize{68--566} & \scriptsize{314} & \scriptsize{\nodata} \\
\scriptsize{{\hi} (Br19)} & \scriptsize{15264.708} & \scriptsize{\nodata} & \scriptsize{\nodata} & \scriptsize{12.749} & \scriptsize{13.561} & \scriptsize{$-1.414$} & \scriptsize{\nodata} & \scriptsize{\nodata}  & \scriptsize{\nodata} & \scriptsize{\nodata} & \scriptsize{DIB~15271} \\
\scriptsize{Fe~I} & \scriptsize{15298.740} & \scriptsize{15298.528} & \scriptsize{$0.212$} & \scriptsize{5.309} & \scriptsize{6.119} & \scriptsize{$0.650$} & \scriptsize{6} & \scriptsize{3}  & \scriptsize{188--283} & \scriptsize{228} & \scriptsize{\nodata} \\
\scriptsize{{\hi} (Br18)} & \scriptsize{15345.982} & \scriptsize{15345.987} & \scriptsize{$-0.005$} & \scriptsize{12.749} & \scriptsize{13.556} & \scriptsize{$-1.337$} & \scriptsize{\nodata} & \scriptsize{57}  & \scriptsize{67--533} & \scriptsize{276} & \scriptsize{\nodata} \\
\scriptsize{{\hi} (Br17)} & \scriptsize{15443.139} & \scriptsize{15443.187} & \scriptsize{$-0.048$} & \scriptsize{12.749} & \scriptsize{13.551} & \scriptsize{$-1.255$} & \scriptsize{\nodata} & \scriptsize{73}  & \scriptsize{66--537} & \scriptsize{266} & \scriptsize{\nodata} \\
\scriptsize{{\hi} (Br16)} & \scriptsize{15560.699} & \scriptsize{15560.697} & \scriptsize{$0.002$} & \scriptsize{12.749} & \scriptsize{13.545} & \scriptsize{$-1.166$} & \scriptsize{\nodata} & \scriptsize{47}  & \scriptsize{65--436} & \scriptsize{266} & \scriptsize{\nodata} \\
\scriptsize{N~I} & \scriptsize{15586.545} & \scriptsize{15586.591} & \scriptsize{$-0.046$} & \scriptsize{12.126} & \scriptsize{12.922} & \scriptsize{$-0.023$} & \scriptsize{1} & \scriptsize{1}  & \scriptsize{52--52} & \scriptsize{52} & \scriptsize{\nodata} \\
\scriptsize{{\hi} (Br15)} & \scriptsize{15704.952} & \scriptsize{15705.015} & \scriptsize{$-0.063$} & \scriptsize{12.749} & \scriptsize{13.538} & \scriptsize{$-1.071$} & \scriptsize{\nodata} & \scriptsize{76}  & \scriptsize{65--552} & \scriptsize{282} & \scriptsize{\nodata} \\
\scriptsize{Mg~I} & \scriptsize{15753.291} & \scriptsize{blend} & \scriptsize{\nodata} & \scriptsize{5.932} & \scriptsize{6.719} & \scriptsize{$0.140$} & \scriptsize{7} & \scriptsize{\nodata} & \scriptsize{\nodata} & \scriptsize{\nodata} & \scriptsize{{Mg~{\sc i}}~15745} \\
\scriptsize{$\lambda15760$} & \scriptsize{\nodata} & \scriptsize{15760.161} & \scriptsize{\nodata} & \scriptsize{\nodata} & \scriptsize{\nodata} & \scriptsize{\nodata} & \scriptsize{36 (13)} & \scriptsize{11}  & \scriptsize{27--304} & \scriptsize{161} & \scriptsize{{\mgi}} \\
\scriptsize{Mg~I} & \scriptsize{15770.149} & \scriptsize{15770.943} & \scriptsize{$-0.794$} & \scriptsize{5.933} & \scriptsize{6.719} & \scriptsize{$0.411$} & \scriptsize{9 (1)} & \scriptsize{2}  & \scriptsize{338--375} & \scriptsize{356} & \scriptsize{$\lambda15760$} \\
\scriptsize{{\hi} (Br14)} & \scriptsize{15884.880} & \scriptsize{15884.875} & \scriptsize{$0.005$} & \scriptsize{12.749} & \scriptsize{13.529} & \scriptsize{$-0.967$} & \scriptsize{\nodata} & \scriptsize{91}  & \scriptsize{61--522} & \scriptsize{259} & \scriptsize{\nodata} \\
\scriptsize{Si~I} & \scriptsize{15892.771} & \scriptsize{blend} & \scriptsize{\nodata} & \scriptsize{5.082} & \scriptsize{5.862} & \scriptsize{$-0.007$} & \scriptsize{6} & \scriptsize{\nodata} & \scriptsize{\nodata} & \scriptsize{\nodata} & \scriptsize{\nodata} \\
\scriptsize{Si~I} & \scriptsize{15964.422} & \scriptsize{15963.229} & \scriptsize{$1.193$} & \scriptsize{5.984} & \scriptsize{6.761} & \scriptsize{$0.198$} & \scriptsize{7 (1)} & \scriptsize{3}  & \scriptsize{321--383} & \scriptsize{345} & \scriptsize{\nodata} \\
\scriptsize{C~I} & \scriptsize{16009.270} & \scriptsize{blend} & \scriptsize{\nodata} & \scriptsize{9.631} & \scriptsize{10.406} & \scriptsize{$0.234$} & \scriptsize{7 (3)} & \scriptsize{\nodata} & \scriptsize{\nodata} & \scriptsize{\nodata} & \scriptsize{\nodata} \\
\scriptsize{C~I} & \scriptsize{16026.080} & \scriptsize{blend} & \scriptsize{\nodata} & \scriptsize{9.631} & \scriptsize{10.405} & \scriptsize{$0.222$} & \scriptsize{5 (5)} & \scriptsize{\nodata} & \scriptsize{\nodata} & \scriptsize{\nodata} & \scriptsize{\nodata} \\
\scriptsize{{\hi} (Br13)} & \scriptsize{16113.714} & \scriptsize{16113.766} & \scriptsize{$-0.052$} & \scriptsize{12.749} & \scriptsize{13.518} & \scriptsize{$-0.852$} & \scriptsize{\nodata} & \scriptsize{96}  & \scriptsize{61--517} & \scriptsize{249} & \scriptsize{\nodata} \\
\scriptsize{{\hi} (Br12)} & \scriptsize{16411.674} & \scriptsize{16411.763} & \scriptsize{$-0.089$} & \scriptsize{12.749} & \scriptsize{13.504} & \scriptsize{$-0.725$} & \scriptsize{\nodata} & \scriptsize{95}  & \scriptsize{59--524} & \scriptsize{252} & \scriptsize{\nodata} \\
\scriptsize{Ca~II} & \scriptsize{16565.590} & \scriptsize{16565.973} & \scriptsize{$-0.383$} & \scriptsize{9.235} & \scriptsize{9.983} & \scriptsize{$0.368$} & \scriptsize{6} & \scriptsize{3}  & \scriptsize{208--333} & \scriptsize{291} & \scriptsize{{C~{\sc i}}~16564} \\
\scriptsize{Ca~II} & \scriptsize{16654.430} & \scriptsize{blend} & \scriptsize{\nodata} & \scriptsize{9.240} & \scriptsize{9.984} & \scriptsize{$0.626$} & \scriptsize{1 (2)} & \scriptsize{\nodata} & \scriptsize{\nodata} & \scriptsize{\nodata} & \scriptsize{\nodata} \\
\scriptsize{Si~I} & \scriptsize{16685.341} & \scriptsize{blend} & \scriptsize{\nodata} & \scriptsize{5.984} & \scriptsize{6.727} & \scriptsize{$-0.117$} & \scriptsize{1 (2)} & \scriptsize{\nodata} & \scriptsize{\nodata} & \scriptsize{\nodata} & \scriptsize{\nodata} \\
\scriptsize{Mg~II} & \scriptsize{16764.800} & \scriptsize{16764.922} & \scriptsize{$-0.122$} & \scriptsize{12.083} & \scriptsize{12.822} & \scriptsize{$0.481$} & \scriptsize{2 (2)} & \scriptsize{1}  & \scriptsize{34--34} & \scriptsize{34} & \scriptsize{\nodata} \\
\scriptsize{$\lambda16781$} & \scriptsize{\nodata} & \scriptsize{16781.115} & \scriptsize{\nodata} & \scriptsize{\nodata} & \scriptsize{\nodata} & \scriptsize{\nodata} & \scriptsize{36 (13)} & \scriptsize{13}  & \scriptsize{28--309} & \scriptsize{157} & \scriptsize{\nodata} \\
\scriptsize{Fe~II} & \scriptsize{16791.762} & \scriptsize{16791.953} & \scriptsize{$-0.191$} & \scriptsize{5.484} & \scriptsize{6.222} & \scriptsize{$-2.325$} & \scriptsize{8 (6)} & \scriptsize{1}  & \scriptsize{48--48} & \scriptsize{48} & \scriptsize{\nodata} \\
\scriptsize{Mg~II} & \scriptsize{16804.520} & \scriptsize{blend} & \scriptsize{\nodata} & \scriptsize{12.085} & \scriptsize{12.822} & \scriptsize{$0.737$} & \scriptsize{2 (2)} & \scriptsize{\nodata} & \scriptsize{\nodata} & \scriptsize{\nodata} & \scriptsize{{Mg~{\sc ii}}~16804} \\
\scriptsize{{\hi} (Br11)} & \scriptsize{16811.111} & \scriptsize{\nodata} & \scriptsize{\nodata} & \scriptsize{12.749} & \scriptsize{13.486} & \scriptsize{$-0.582$} & \scriptsize{\nodata} & \scriptsize{194}  & \scriptsize{57--1153} & \scriptsize{282} & \scriptsize{\nodata} \\
\scriptsize{Fe~II} & \scriptsize{16877.808} & \scriptsize{16877.822} & \scriptsize{$-0.014$} & \scriptsize{5.484} & \scriptsize{6.219} & \scriptsize{$-1.256$} & \scriptsize{76 (33)} & \scriptsize{26}  & \scriptsize{24--232} & \scriptsize{82} & \scriptsize{\nodata} \\
\scriptsize{C~I} & \scriptsize{16895.031} & \scriptsize{16894.898} & \scriptsize{$0.133$} & \scriptsize{9.003} & \scriptsize{9.736} & \scriptsize{$0.534$} & \scriptsize{43 (19)} & \scriptsize{14}  & \scriptsize{31--539} & \scriptsize{243} & \scriptsize{\nodata} \\
\scriptsize{Si~II} & \scriptsize{16911.430} & \scriptsize{16911.646} & \scriptsize{$-0.216$} & \scriptsize{12.147} & \scriptsize{12.880} & \scriptsize{$0.350$} & \scriptsize{1 (1)} & \scriptsize{1}  & \scriptsize{41--41} & \scriptsize{41} & \scriptsize{\nodata} \\
\enddata
\tablenotetext{1}{\scriptsize{The emission peak midpoint corrected by the emission peak midpoint of Br11. }}
\end{deluxetable}

\newpage
\clearpage
\begin{deluxetable}{lcclccr}
\tablecolumns{7}
\tablecaption{Line-emitting Disk Radius Estimates \label{table_rdisk}}
\tablehead{\colhead{\scriptsize{ABE}} & \colhead{\scriptsize{Lit.}}     & \colhead{\scriptsize{Ref.}} & \colhead{\scriptsize{Atom}} & \colhead{\scriptsize{\#}}  & \colhead{\scriptsize{Mean}}     & \colhead{\scriptsize{Mean}} \\
           \colhead{\scriptsize{ID}}  & \colhead{\scriptsize{{\vsini}}} & \colhead{}                  & \colhead{\scriptsize{or}}   & \colhead{\tiny{spectra}}   & \colhead{\scriptsize{{\vp}}}    & \colhead{\scriptsize{{\rp}}}  \\
           \colhead{}                 & \colhead{\scriptsize{[{\kms}]}} & \colhead{}                  & \colhead{\scriptsize{Ion}}  & \colhead{}                 & \colhead{\scriptsize{[{\kms}]}} & \colhead{\scriptsize{[{\rs}]}}  }
\startdata
\scriptsize{001\tablenotemark{a}} & \scriptsize{$266$} & \scriptsize{9} & \scriptsize{Br11} & \scriptsize{4} & \scriptsize{$208$} & \scriptsize{6.56} \\
\scriptsize{} & \scriptsize{} & \scriptsize{} & \scriptsize{$\lambda15760$} & \scriptsize{4} & \scriptsize{$246$} & \scriptsize{4.67} \\
\scriptsize{} & \scriptsize{} & \scriptsize{} & \scriptsize{$\lambda16781$} & \scriptsize{4} & \scriptsize{$261$} & \scriptsize{4.16} \\
\scriptsize{003} & \scriptsize{$225$} & \scriptsize{5} & \scriptsize{Br11} & \scriptsize{10} & \scriptsize{$305$} & \scriptsize{2.18} \\
\scriptsize{006} & \scriptsize{$231$} & \scriptsize{7} & \scriptsize{Br11} & \scriptsize{3} & \scriptsize{$294$} & \scriptsize{2.47} \\
\scriptsize{014} & \scriptsize{$230$} & \scriptsize{3} & \scriptsize{Br11} & \scriptsize{7} & \scriptsize{$326$} & \scriptsize{1.99} \\
\scriptsize{016\tablenotemark{a}} & \scriptsize{$250$} & \scriptsize{4} & \scriptsize{Br11} & \scriptsize{1} & \scriptsize{$238$} & \scriptsize{4.43} \\
\scriptsize{} & \scriptsize{} & \scriptsize{} & \scriptsize{{\feii}~16878} & \scriptsize{1} & \scriptsize{$111$} & \scriptsize{20.41} \\
\scriptsize{026\tablenotemark{b}} & \scriptsize{$230$} & \scriptsize{3} & \scriptsize{Br11} & \scriptsize{12} & \scriptsize{$324$} & \scriptsize{2.02} \\
\scriptsize{028} & \scriptsize{$242$} & \scriptsize{4} & \scriptsize{Br11} & \scriptsize{12} & \scriptsize{$297$} & \scriptsize{2.66} \\
\scriptsize{036\tablenotemark{b}} & \scriptsize{$307$} & \scriptsize{4} & \scriptsize{Br11} & \scriptsize{3} & \scriptsize{$456$} & \scriptsize{1.81} \\
\scriptsize{046\tablenotemark{b}} & \scriptsize{$235$} & \scriptsize{2} & \scriptsize{Br11} & \scriptsize{1} & \scriptsize{$294$} & \scriptsize{2.56} \\
\scriptsize{049} & \scriptsize{$120$} & \scriptsize{2} & \scriptsize{Br11} & \scriptsize{23} & \scriptsize{$197$} & \scriptsize{1.48} \\
\scriptsize{} & \scriptsize{} & \scriptsize{} & \scriptsize{{\feii}~16878} & \scriptsize{7} & \scriptsize{$67$} & \scriptsize{12.73} \\
\scriptsize{067} & \scriptsize{$120$} & \scriptsize{5} & \scriptsize{Br11} & \scriptsize{8} & \scriptsize{$133$} & \scriptsize{3.24} \\
\scriptsize{} & \scriptsize{} & \scriptsize{} & \scriptsize{{\feii}~16878} & \scriptsize{8} & \scriptsize{$41$} & \scriptsize{34.32} \\
\scriptsize{085} & \scriptsize{$182$} & \scriptsize{8} & \scriptsize{Br11} & \scriptsize{4} & \scriptsize{$321$} & \scriptsize{1.29} \\
\scriptsize{098} & \scriptsize{$166$} & \scriptsize{11} & \scriptsize{Br11} & \scriptsize{1} & \scriptsize{$301$} & \scriptsize{1.21} \\
\scriptsize{133\tablenotemark{b}} & \scriptsize{$286$} & \scriptsize{4} & \scriptsize{Br11} & \scriptsize{3} & \scriptsize{$285$} & \scriptsize{4.02} \\
\scriptsize{138} & \scriptsize{$243$} & \scriptsize{7} & \scriptsize{Br11} & \scriptsize{2} & \scriptsize{$365$} & \scriptsize{1.77} \\
\scriptsize{140} & \scriptsize{$328$} & \scriptsize{4} & \scriptsize{Br11} & \scriptsize{4} & \scriptsize{$382$} & \scriptsize{2.95} \\
\scriptsize{165} & \scriptsize{$274$} & \scriptsize{4} & \scriptsize{Br11} & \scriptsize{3} & \scriptsize{$397$} & \scriptsize{1.91} \\
\scriptsize{167} & \scriptsize{$160$} & \scriptsize{7} & \scriptsize{Br11} & \scriptsize{3} & \scriptsize{$300$} & \scriptsize{1.14} \\
\scriptsize{170} & \scriptsize{$130$} & \scriptsize{5} & \scriptsize{Br11} & \scriptsize{6} & \scriptsize{$153$} & \scriptsize{2.90} \\
\scriptsize{} & \scriptsize{} & \scriptsize{} & \scriptsize{{\feii}~16878} & \scriptsize{6} & \scriptsize{$64$} & \scriptsize{16.76} \\
\scriptsize{} & \scriptsize{} & \scriptsize{} & \scriptsize{{\ci}~16895} & \scriptsize{2} & \scriptsize{$169$} & \scriptsize{2.36} \\
\scriptsize{189} & \scriptsize{$256$} & \scriptsize{6} & \scriptsize{Br11} & \scriptsize{5} & \scriptsize{$305$} & \scriptsize{2.81} \\
\scriptsize{191} & \scriptsize{$242$} & \scriptsize{4} & \scriptsize{Br11} & \scriptsize{1} & \scriptsize{$274$} & \scriptsize{3.11} \\
\scriptsize{195\tablenotemark{b}} & \scriptsize{$148$} & \scriptsize{7} & \scriptsize{Br11} & \scriptsize{3} & \scriptsize{$266$} & \scriptsize{1.24} \\
\scriptsize{A08} & \scriptsize{$343$} & \scriptsize{11} & \scriptsize{Br11} & \scriptsize{3} & \scriptsize{$409$} & \scriptsize{2.82} \\
\scriptsize{A11} & \scriptsize{$220$} & \scriptsize{7} & \scriptsize{Br11} & \scriptsize{3} & \scriptsize{$215$} & \scriptsize{4.19} \\
\scriptsize{A12} & \scriptsize{$215$} & \scriptsize{7} & \scriptsize{Br11} & \scriptsize{1} & \scriptsize{$358$} & \scriptsize{1.44} \\
\scriptsize{A13\tablenotemark{a}\tablenotemark{b}} & \scriptsize{$306$} & \scriptsize{4} & \scriptsize{Br11} & \scriptsize{6} & \scriptsize{$257$} & \scriptsize{5.68} \\
\scriptsize{} & \scriptsize{} & \scriptsize{} & \scriptsize{$\lambda15760$} & \scriptsize{4} & \scriptsize{$248$} & \scriptsize{6.07} \\
\scriptsize{} & \scriptsize{} & \scriptsize{} & \scriptsize{$\lambda16781$} & \scriptsize{4} & \scriptsize{$250$} & \scriptsize{6.00} \\
\scriptsize{A15\tablenotemark{a}} & \scriptsize{$350$} & \scriptsize{7} & \scriptsize{Br11} & \scriptsize{3} & \scriptsize{$237$} & \scriptsize{8.76} \\
\scriptsize{} & \scriptsize{} & \scriptsize{} & \scriptsize{{\feii}~16878} & \scriptsize{3} & \scriptsize{$232$} & \scriptsize{9.07} \\
\scriptsize{} & \scriptsize{} & \scriptsize{} & \scriptsize{{\ci}~16895} & \scriptsize{3} & \scriptsize{$345$} & \scriptsize{4.12} \\
\scriptsize{} & \scriptsize{} & \scriptsize{} & \scriptsize{$\lambda15760$} & \scriptsize{3} & \scriptsize{$233$} & \scriptsize{9.05} \\
\scriptsize{} & \scriptsize{} & \scriptsize{} & \scriptsize{$\lambda16781$} & \scriptsize{3} & \scriptsize{$233$} & \scriptsize{9.04} \\
\scriptsize{A17\tablenotemark{b}} & \scriptsize{$300$} & \scriptsize{1} & \scriptsize{Br11} & \scriptsize{3} & \scriptsize{$241$} & \scriptsize{6.19} \\
\scriptsize{} & \scriptsize{} & \scriptsize{} & \scriptsize{{\feii}~16878} & \scriptsize{3} & \scriptsize{$102$} & \scriptsize{34.32} \\
\scriptsize{A20} & \scriptsize{$210$} & \scriptsize{4} & \scriptsize{Br11} & \scriptsize{3} & \scriptsize{$358$} & \scriptsize{1.37} \\
\scriptsize{A29\tablenotemark{a}} & \scriptsize{$222$} & \scriptsize{4} & \scriptsize{Br11} & \scriptsize{3} & \scriptsize{$197$} & \scriptsize{5.09} \\
\scriptsize{} & \scriptsize{} & \scriptsize{} & \scriptsize{{\feii}~16878} & \scriptsize{3} & \scriptsize{$121$} & \scriptsize{13.42} \\
\scriptsize{} & \scriptsize{} & \scriptsize{} & \scriptsize{$\lambda15760$} & \scriptsize{3} & \scriptsize{$144$} & \scriptsize{9.55} \\
\scriptsize{} & \scriptsize{} & \scriptsize{} & \scriptsize{$\lambda16781$} & \scriptsize{3} & \scriptsize{$144$} & \scriptsize{9.52} \\
\scriptsize{A32} & \scriptsize{$260$} & \scriptsize{10} & \scriptsize{Br11} & \scriptsize{3} & \scriptsize{$380$} & \scriptsize{1.87} \\
\enddata
\tablenotetext{1}{Asymmetric emission peak intensities.}
\tablenotetext{2}{Shell stars.}
\tablerefs{\scriptsize{[1] \citet{1970crvs.book.....U}; [2] \citet{1982csrv.book.....U}; [3] \citet{1996PASP..108..833H}; [4] \citet{2001A&A...368..912Y}; [5] \citet{2002ApJ...573..359A}; [6] \citet{2005A&A...440..305F}; [7] \citet{2006A&A...451.1053F}; [8] \citet{2006ApJ...648..580H}; [9] \citet{2007BASI...35..383B}; [10] \citet{2010A&A...522A.107R}; [11] \citet{2010ApJ...722..605H}}}
 
\end{deluxetable}

\clearpage

\large{\textbf{Supplemental data table}}

\vspace{1 cm}An expanded version of Table~\ref{table_stars} is provided below, including ABE identifier, 2MASS designation, HD number, an alternate identifier, $V$ magnitudes from SIMBAD, $H$ magnitudes from 2MASS, the spectral type and reference, and finally, line detection or {\vp} measures for Br11 and the stronger metallic lines. If a {\vp} measurement could not be made in any of the available spectra despite evidence of emission or shell absorption in a line, one of the following abbreviations is provided in place of a {\vp} value: ``\textbf{w}'' detection but lack of visible emission peaks, ``\textbf{w?}'' weak or ambiguous detection, ``\textbf{sp}'' single-peaked emission, ``\textbf{sh}'' shell absorption without resolved adjacent emission peaks, ``\textbf{as}'' highly asymmetric emission peaks such that only one is visible (ABE-056 and ABE-A03), ``\textbf{bl}'' V peak of Br11 is severely blended with {\feii}~16792 (ABE-013), ``\textbf{tc}'' spectra are heavily contaminated by telluric features (ABE-058).

\begin{longtable}{llclcrrclccccc}
\caption{\label{table_stars_big} List of ABE stars}\\
\hline\hline
\tiny{ABe} &  \tiny{2MASS ID} &  \tiny{HD} & \tiny{Other} & \tiny{V} & \tiny{H} & \tiny{Spec Type} & \tiny{Ref.} &
 &  & \tiny{{\vp}} &  &  &  \\
 &  &  &  &  &  &  &  &
\tiny{Br11} & \tiny{$\lambda15760$} & \tiny{$\lambda16781$} & \tiny{{\feii}~16792} & \tiny{{\feii}~16878} & \tiny{{\ci}~16895} \\
 &  &  &  & \tiny{mag} & \tiny{mag} &  &  &
\tiny{km s$^{-1}$} & \tiny{km s$^{-1}$} & \tiny{km s$^{-1}$} & \tiny{km s$^{-1}$} & \tiny{km s$^{-1}$} & \tiny{km s$^{-1}$} \\
\hline
\endfirsthead
\caption{continued.}\\
\hline\hline
\tiny{ABe} &  \tiny{2MASS ID} &  \tiny{HD} & \tiny{Other} & \tiny{V} & \tiny{H} & \tiny{Spec Type} & \tiny{Ref.} &
 &  & \tiny{{\vp}} &  &  &  \\
 &  &  &  &  &  &  &  &
\tiny{Br11} & \tiny{$\lambda15760$} & \tiny{$\lambda16781$} & \tiny{{\feii}~16792} & \tiny{{\feii}~16878} & \tiny{{\ci}~16895} \\
 &  &  &  & \tiny{mag} & \tiny{mag} &  &  &
\tiny{km s$^{-1}$} & \tiny{km s$^{-1}$} & \tiny{km s$^{-1}$} & \tiny{km s$^{-1}$} & \tiny{km s$^{-1}$} & \tiny{km s$^{-1}$} \\
\hline
\endhead
\hline
\endfoot

\tiny{001} & \tiny{$20212485+3722482$} & \tiny{\nodata} & \tiny{VES 203} & \tiny{12.09} & \tiny{9.108} & \tiny{B0.5Ve} & \tiny{74} & \tiny{208} & \tiny{246} & \tiny{261} & \tiny{-} & \tiny{w} & \tiny{-} \\
\tiny{002} & \tiny{$20151525+3654562$} & \tiny{228576} & \tiny{AS 394} & \tiny{11.41} & \tiny{9.888} & \tiny{Ae} & \tiny{29} & \tiny{152} & \tiny{w} & \tiny{86} & \tiny{-} & \tiny{85} & \tiny{-} \\
\tiny{003} & \tiny{$20162816+3703229$} & \tiny{192987} & \tiny{HR 7757} & \tiny{6.46} & \tiny{6.548} & \tiny{B6IIIe} & \tiny{22} & \tiny{305} & \tiny{-} & \tiny{-} & \tiny{-} & \tiny{w?} & \tiny{w} \\
\tiny{004} & \tiny{$20234596+3830033$} & \tiny{229221} & \tiny{MWC 344} & \tiny{9.22} & \tiny{6.734} & \tiny{B0.2IIIe} & \tiny{66} & \tiny{115} & \tiny{131} & \tiny{105} & \tiny{-} & \tiny{w?} & \tiny{191} \\
\tiny{005} & \tiny{$20184170+3759106$} & \tiny{\nodata} & \tiny{Hen 3-1876} & \tiny{11.37} & \tiny{9.699} & \tiny{OB} & \tiny{6} & \tiny{271} & \tiny{w} & \tiny{w} & \tiny{-} & \tiny{-} & \tiny{-} \\
\tiny{006} & \tiny{$19124025-0627316$} & \tiny{179405} & \tiny{MWC 615} & \tiny{9.12} & \tiny{8.084} & \tiny{B2Ve} & \tiny{68} & \tiny{294} & \tiny{-} & \tiny{-} & \tiny{-} & \tiny{-} & \tiny{-} \\
\tiny{007\tablenotemark{new}} & \tiny{$19104149-0542581$} & \tiny{178920} & \tiny{BD-05 4897} & \tiny{10.38} & \tiny{9.236} & \tiny{B8II/III} & \tiny{58} & \tiny{181} & \tiny{-} & \tiny{-} & \tiny{-} & \tiny{w} & \tiny{w?} \\
\tiny{008\tablenotemark{new}} & \tiny{$18000176-2323071$} & \tiny{\nodata} & \tiny{\nodata} & \tiny{12.20} & \tiny{10.699} & \tiny{\nodata} & \tiny{\nodata} & \tiny{233} & \tiny{-} & \tiny{-} & \tiny{-} & \tiny{w} & \tiny{w?} \\
\end{longtable}

\newpage
\clearpage

\newpage
\clearpage

\newpage
\clearpage

\newpage
\clearpage

\newpage
\clearpage

\newpage
\clearpage

\newpage
\clearpage

\newpage
\clearpage

\newpage
\clearpage

\newpage
\clearpage

\newpage
\clearpage

\newpage
\clearpage

\newpage
\clearpage

\newpage
\clearpage

\newpage
\clearpage

\newpage
\clearpage

\newpage
\clearpage

\newpage
\clearpage

\newpage
\clearpage

\clearpage

\appendix \label{atlas}
\section{} 
\large{\textbf{Supplemental figures}}
\normalsize

\vspace{0.5 cm}
\begin{figure*}[h!]
\includegraphics[width=18cm]{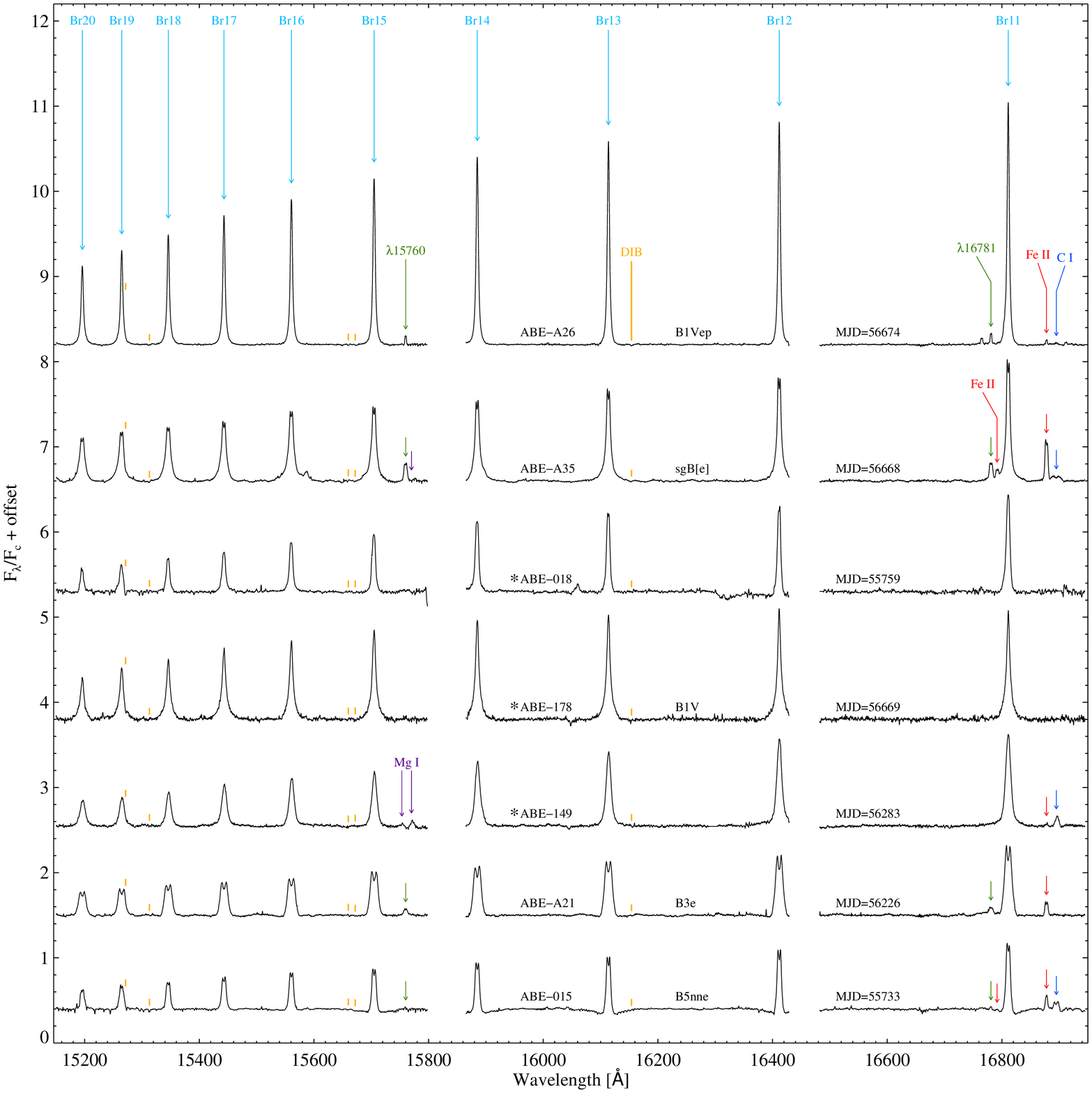}
\caption{Full APOGEE spectra for Be stars with strong Brackett series emission. The emission lines for ABE-A35, ABE-A21, and ABE-015 are double-peaked, whereas the other stars have single-peaked emission. ABE identifiers, observation MJDs and literature spectral types for each star are provided, and newly-identified-via-APOGEE Be stars are indicated with a large asterisk preceding the ABE identifier. Small line segments mark the positions of the most prominent DIBs appearing for these stars, and arrows mark the positions of the most frequently-detected metallic emission features. \label{fig_atlas1}}
\end{figure*}

\begin{figure*}
\includegraphics[width=18cm]{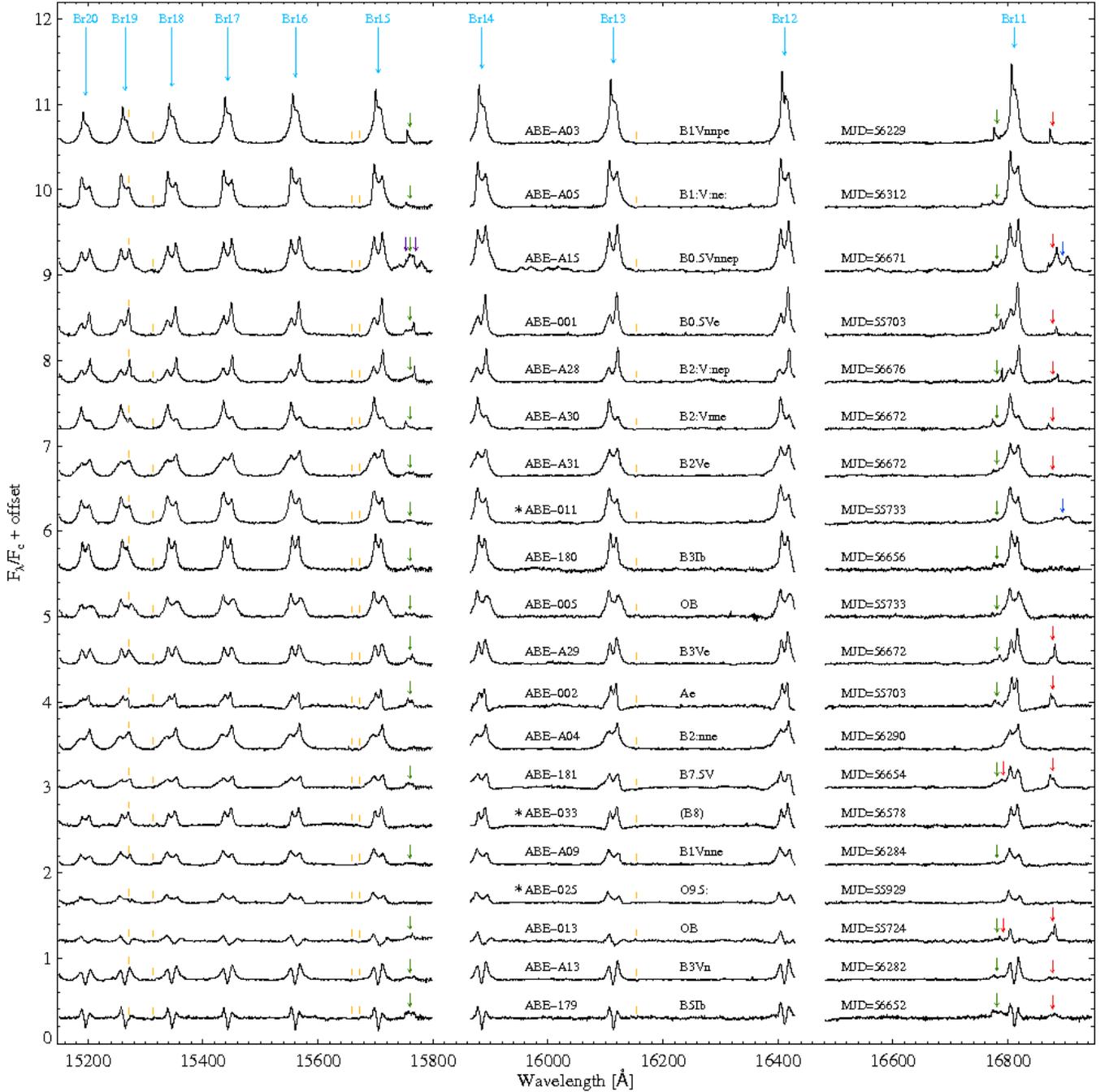}
\caption{Full APOGEE spectra for a selection of 20 Be stars with asymmetric Brackett series emission. Note the striking similarity between the spectra of ABE-001, ABE-A28, and ABE-A30 (the latter a V/R reflection of the former two). Meanings are the same as in Figure~\ref{fig_atlas1}. \label{fig_atlas2}}
\end{figure*}

\begin{figure*}
\includegraphics[width=18cm]{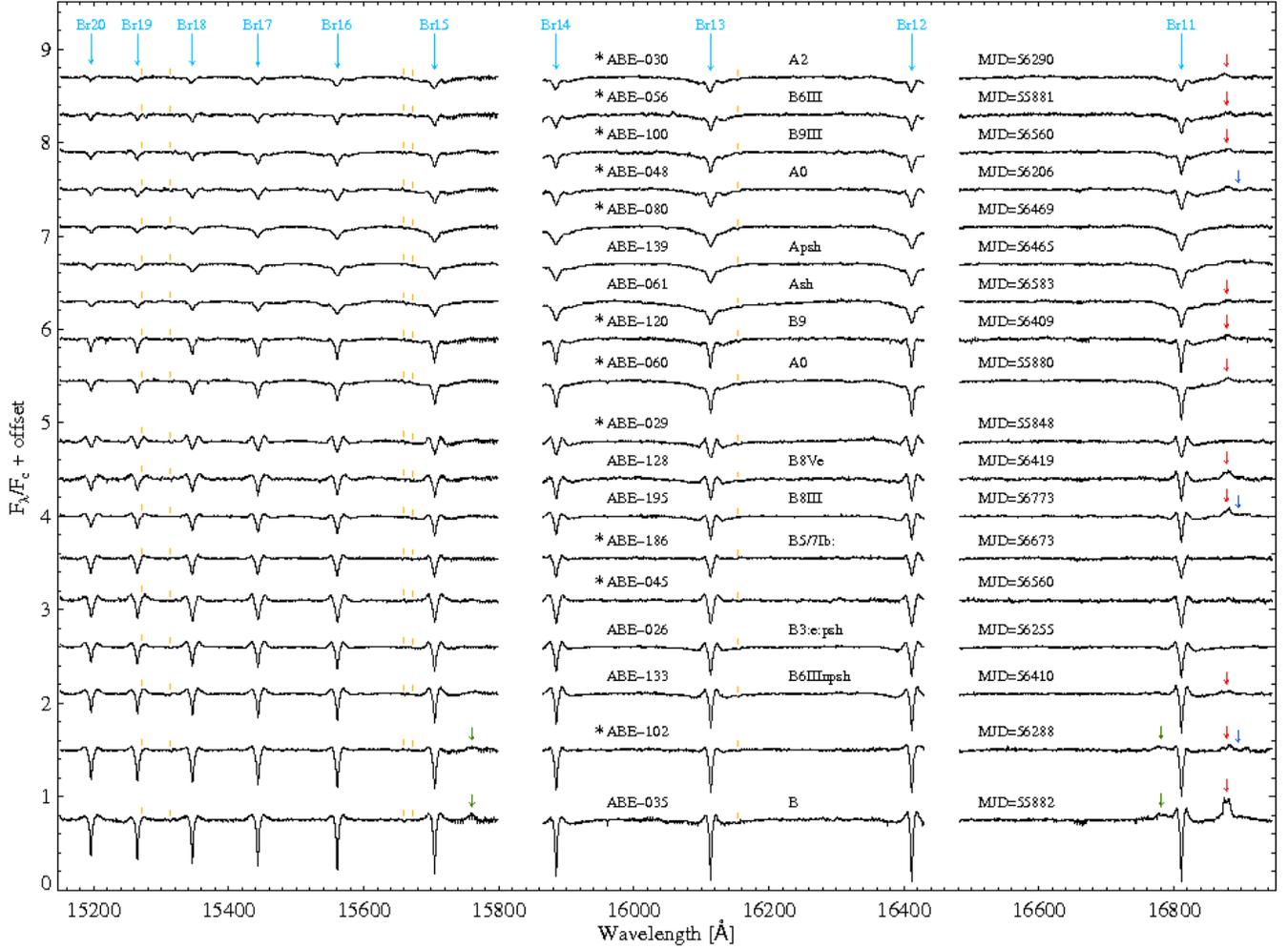}
\caption{Full APOGEE spectra for a selection of 18 Be-shell stars. Broad photospheric absorption wings are clearly evident in the Brackett lines for the upper 9 stars, while the lower 9 stars exhibit mostly smooth continua and shell features with adjacent emission. Meanings are the same as in Figure~\ref{fig_atlas1}. \label{fig_atlas2}}
\end{figure*}

\end{document}